%% file: multiple_protocols.tex
\documentclass[10pt, onecolumn]{IEEEtran}

\usepackage{geometry}
\usepackage{multicol,graphicx,amssymb,mathrsfs,amsmath,pifont,amscd,latexsym,amsthm,color,fancyhdr,CJK,url,hyperref,longtable, pifont, subfigure,epstopdf,setspace,multirow,colortbl,tabularx,threeparttable, booktabs, verbatim, bbm,listings, balance}
\usepackage[linesnumbered,boxed,algosection]{algorithm2e}
\definecolor{mygray}{gray}{.7}

\begin{document}

\title{Demystifying Mobile Web Browsing under Multiple Protocols}
\author{Yi Liu}

\maketitle

\begin{abstract}
With the popularity of mobile devices, such as smartphones, tablets, users prefer visiting Web pages on mobile devices. Meanwhile, HTTP(S) plays as the major protocol to deliver Web contents, and has served the Web well for more than 15 years. However, as the Web pages grow increasingly complex to provide more content and functionality, the shortcomings and inflexibility of HTTP become more and more urgent to solve~\cite{Stenberg:CCR14}~\cite{httpproblem}, e.g., the sluggish page load, insecure content, redundant transfer, etc. SPDY and HTTP/2 are promoted to solve the shortcomings and inflexibilities of HTTP/1.x. We are interested in how Web pages perform on smartphones with different protocols, including HTTP, HTTPS, SPDY, and HTTP/2. In this paper, we divide our experiments into two parts. First, in order to simplify our analysis, we develop our own HTTP client ignoring complicated process in real browsers to fetch synthetic Web pages with pre-specified object sizes and object numbers with different protocols, respectively. Meanwhile, we emulate different network conditions between client and server using Traffic Control. In order to test with real browsers, we clone Alexa top 200 websites, which have the corresponding mobile version, into our local host. Meanwhile, we control mobile Chrome browser to load those Web pages with different protocols and emulate different network conditions using Traffic Control. We identify how Web page characteristics and network conditions affect Web performance on smartphones for each protocol. We also conduct experiments on a low-end device to observe if a less powerful processor could affect Web page performance for each protocol.
\end{abstract}

\input{section/introduction}
\input{section/background}
\input{section/measurement_setup}
\input{section/performance_protocols}

\input{section/web_page_performance}
\input{section/how_http2_help}
\input{section/findings_and_implications}
\input{section/relative_work}
\input{section/conclusion}

\bibliography{ref}
\bibliographystyle{IEEEtran}

\end{document}

%% file: section/introduction.tex
\section{Introduction}
% no \IEEEPARstart

HTTP(S) plays as the major protocol to access Web contents when loading Web pages on browsers, and has served the Web well for more than 15 years. However, as the Web pages grow increasingly complex to provide more content and functionality, the shortcomings and inflexibly of HTTP become more and more urgent to solve~\cite{Stenberg:CCR14}~\cite{httpproblem}, e.g., the sluggish page load, insecure content, redundant transfer, etc. HTTP/2 is the next evolution of HTTP, which is maintained by the IETF HTTP Working Group. Based on Google's SPDY protocol, HTTP/2 attempts to outcome the shortcomings of HTTP, and focuses on performance, e.g., end-user perceived latency, network, and server resource usage. HTTP/2 and SPDY benefit from multiplexing and concurrency, stream dependencies, header compressions, and server push~\cite{Stenberg:CCR14}. All the four protocols, including HTTP, HTTPS, SPDY, and HTTP/2 can be affected by many factors external to the protocols, including the network parameters (e.g. packet loss, bandwidth, and RTT), and the characteristics of Web pages (e.g. size, objects number)~\cite{Wang:NSDI14,Saxce:INFCOM15, Varvello:CORR15, Erman:CONEXT13, El-khatibTW:IFIP14}. Meanwhile, with the popularity of mobile devices such as smartphones and tablets, users often use mobile browsers to visit Web pages, including both desktop versions and mobile versions. Such mobile pages are optimized for mobile devices with limited resources and smaller screen. Although WAP-based Web technologies are developed for feature phones, HTTP(S) plays as the major protocol to access Web contents for contemporary smartphones. We just focus on the HTTP-based protocols, and we are interested in how Web pages perform on smartphones with different protocols, including HTTP, HTTPS, SPDY, and HTTP/2.

Although HTTP/2 and SPDY claim to improve Web pages performance instead of HTTP and HTTPS. Prior works~\cite{Wang:NSDI14,Saxce:INFCOM15, Varvello:CORR15, Erman:CONEXT13, El-khatibTW:IFIP14} have shown that they may even hurt Web pages performance under certain network conditions. It is important to conduct a comprehensive study on comparison of Web page load performance with these four protocols. In this paper, we present a comprehensive measurement study to analyze Web performance on smartphones with different protocols. We clone the landing pages of Alexa top 200 websites, which have both mobile version and desktop version, into our local server, and visit them using different protocols through emulated network conditions using \textit{Traffic Control} to keep the consistency across experiments. We vary different packet loss rates, bandwidths, and RTTs (Round Trip Time) to simulate different network conditions and analyze how these network settings affect performance of different protocols. We focus on a user-perceived latency of these pages, so we measure Web page performance using the \textbf{Page Load Time (PLT)} metric. PLT is calculated from initiation phase (when you click a link or type in a Web address on browser) to completion phase (when the page is fully loaded in the browser). We listen the \textit{onLoad}\footnote{Page Load Time metric, \url{https://developer.chrome.com/devtools/docs/network}} event emitted by browser to get the PLT. In order to analyze how different Web pages with various object sizes and object numbers may affect Web pages' performance for each protocol, we synthesize Web pages with pre-specified object sizes and object numbers. Meanwhile, we need to address that loading a Web page in browsers is a complex process, consisting of network activities, including DNS lookup, TCP connection setup, resources downloading and so on, and computation activities, including parsing HTML file, parsing Javascript file, executing Javascript code and so on. What's more, network activities and computation activities are interdependent. For example, while loading a page, browsers usually do not download any images until JavaScript and CSS files are fetched and processed. We develop our own client to fetch Web pages ignoring page load complicated dependencies between network activities and computation activities in order to simplify our analysis. Overall, we identify how those network factors and characteristics of Web pages affect Web page performance for each protocol. 

Our study focuses on performance of Web pages on smartphones with different protocols, including HTTP, HTTPS, SPDY, and HTTP/2. We divide our experiments into two parts. First, in order to simplify our analysis, we develop our own HTTP client ignoring complicated process in real browsers to fetch synthetic Web pages with pre-specified object sizes and object numbers with different protocols, respectively. Meanwhile, we emulate different network conditions between client and server using Traffic Control. We conduct a decision tree analysis to find out how those parameter settings affect these protocols‘ efficiency of transferring Web contents hosted in remote servers. We find that HTTP/2 and SPDY perform worse when packet loss rate is high, but help with many objects when packet loss rate is low. In order to test with real browsers, we clone Alexa top 200 websites, which have the corresponding mobile version, into our local host. Meanwhile, we control mobile Chrome browser to load those Web pages with different protocols and emulate different network conditions using Traffic Control. We identify how Web page characteristics and network conditions affect Web performance on smartphones for each protocol. We find that HTTP/2 and SPDY help more with mobile pages. We also conduct experiments on a low-end device to observe if a less powerful processor could affect Web page performance for each protocol. We find no significant differences for all four protocols under different network conditions.

Our contributions are as follows:

\begin{itemize}
    \item{A systematic and comprehensive measurement study to compare the performance of HTTP, HTTPS, SPDY, and HTTP/2. We identify how different parameter settings could affect Web pages' performance for each protocol. }
    \item{We give some practical implications and recommendations for developers to optimize their pages' performance.}
\end{itemize}

%% file: section/background.tex
\section{Background}
In this section, we demonstrate the shortcomings and inflexibility of HTTP/1.1, and describe the new features and improvement of HTTP/2 and SPDY.

\subsection{Shortcomings and inflexibility of HTTP(S)}

HTTP/1.1 is the defacto standard of HTTP, and has served the Web for more than fifteen years since the standardization. However, the Web has changed a lot to make it outdated. According to HTTP Archive~\cite{httparchive}, a Web page is becoming more and more complex. It may take more than 90 requests over 35 TCP connections to 16 different hosts to load a Web application, which may transfer about 1.9MB data on average. However, HTTP/1.1 practically only allows one outstanding request per TCP connection. Although HTTP/1.1 added the feature of reusing connections, only one request could occupy a TCP channel at a time. Other requests may initiate new TCP connections or wait until the occupied connections released. HTTP/1.1 is very latency sensitive, and has had enough of head of line blocking problem~\cite{headoflineblocking}. 

HTTPS piggybacks HTTP entirely on top of TLS/SSL, so it needs one more expensive TLS/SSL handshake for each new connection, which costs extra two Round-Trip Time (RTT).

Although Web developers have come up with many best practices like domain sharding, spriting, and inlining and concatenation of resources, these techniques have their own shortcommings~\cite{Stenberg:CCR14}. 

\subsection{HTTP/2 and SPDY}

HTTP/2 derives from SPDY, and addresses those shortcomings of HTTP(S). HTTP/2 inherits core features from SPDY~\cite{relationwithSDPY}, so we just present new features of HTTP/2 here.

HTTP/2 uses a single, multiplexed connection, replacing the multiple connections per domain that browsers opened up in HTTP/1.x. HTTP/2 compresses header data and sends it in a concise, binary format, rather than the plain text format used previously. By using a single connection, HTTP/2 reduces the need for several popular HTTP/1.x optimizations, making your Web applications simpler. 

Frame is the smallest unit of communication in HTTP/2, containing a frame header, which at minimum identifies the stream to which the frame belongs. All HTTP/2 communication is performed within a connection that can carry any number of bidirectional streams, each of which is a bidirectional flow of bytes. In turn, each stream communicates in messages, which consist of one or multiple frames, each of which may be interleaved and then reassembled via the embedded stream identifier in the header of each individual frame. There are four key features of HTTP/2:

\begin{itemize}
    \item{\textbf{One connection per origin.} One connection per origin significantly reduces the associated overhead: fewer sockets to manage along the connection path, smaller memory footprint, and better connection throughput. It also may lead to other benefits, such as better compression through use of a single compression context, less time in slow-start, faster congestion and loss recovery.}
    \item{\textbf{Request Prioritization.} HTTP message can be split into many individual frames, the exact order in which the frames are interleaved and delivered can be optimized to further improve the performance. The browser can immediately dispatch each request the moment the resource is discovered, specify the priority of each stream, and let the server determine the optimal response delivery. This eliminates unnecessary request queuing latency and allows us to make the most efficient use of each connection. }
    \item{\textbf{Header Compression.} Each HTTP transfer carries a set of headers that describe the transferred resource and its properties. In HTTP/1.x, this metadata is always sent as plain text and adds anywhere from 500 to 800 bytes of overhead per request, and kilobytes more if HTTP cookies are required. To reduce this overhead and improve performance, HTTP/2 compresses header metadata. Instead of retransmitting the same data on each request and response, HTTP/2 uses ``header table'' on both the client and server to track and store previously sent key-value pairs. Header tables persist for the entire HTTP/2 connection and are incrementally updated both by the client and server. We need to address that SPDY/2 just use a single GZIP context in each direction for header compression.}
    \item{\textbf{Server Push.} When browser loads a Web page in the era of HTTP/1.x, it first initiates a request to fetch the main HTML page. Then, it will initiate consequential requests to fetch other resources, including JavaScript files, CSS files, images when parsing HTML. However, browsers have to initiate a new request for each resource with HTTP/1.x. The key idea of server push is that we could infer further needed resources when loading a page. In additional to the response for the original request, the server can push additional resources to the client. In effect, server push obsoletes most of the cases where inlining is used with HTTP/1.x.}
\end{itemize}

%% file: section/measurement_setup.tex
\section{Measurement Setup}

\begin{figure}[htbp]
	\centering
    \includegraphics[width=0.6\textwidth]{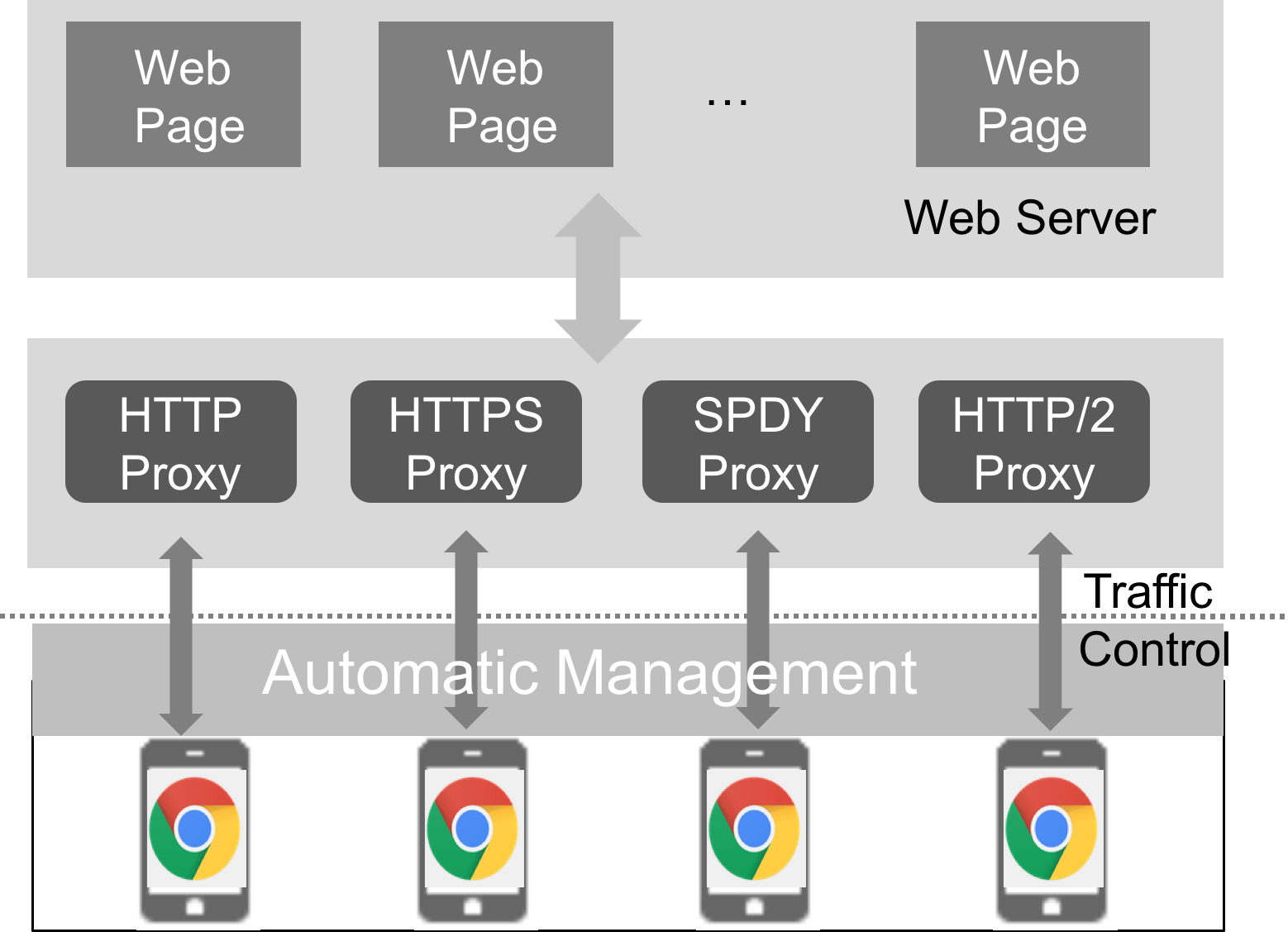}
    \caption{Experiment Setup} \label{fig:experiment_setup}
\end{figure}

In this section, we introduce the infrastructural platforms and how we build the testbed to conduct our experiments as depicted in~\ref{fig:experiment_setup}.

\subsection{Methodology}
We need to address that loading a Web page is not as simple as fetching all resources. For example, while loading a page, browsers usually do not download any images until JavaScript and CSS files are fetched and processed. CSS files may import other CSS files, and the browser can not know about that in advance. Some scripts generate new resources for the browsers to fetch. \cite{Wang:NSDI13} shows that dependencies between network activities (fetching Web objects) and computation activities(HTML parsing, JavaScript execution) have significant impacts on page loading time, so we divide our experiments into two parts. First, we develop our own HTTP client ignoring complicated dependencies between network activities and computation activities in real browsers. We just consider these four protocols as transport protocols to download all resources of Web pages. Then, we conduct our experiments with complete page load dependencies using real browsers to analyze how different protocol perform in real browsing process.

\subsubsection{Experiments as Transport protocols}
 Loading a Web page in browsers is a complex process, consisting of network activities, including DNS lookup, TCP connection setup, resources downloading and so on, and computation activities, including parsing HTML file, parsing Javascript file, executing Javascript code and so on. What's more, network activities and computation activities are interdependent. For example, while loading a page, browsers usually do not download any images until JavaScript and CSS files are fetched and processed. We develop our own client to fetch Web pages ignoring page load complicated dependencies between network activities and computation activities. We just consider these four protocols as transport protocols to simplify our analysis. In order to analyze how different Web pages with various object sizes and object numbers may affect Web pages' performance for each protocol, we synthesize Web pages with pre-specified object sizes and object numbers. Meanwhile, we emulate different network conditions between client and server using \textit{Traffic Control}. We vary different Round Trip Times (RTT), bandwidths, packet loss rates between client and server.
 
 \subsubsection{Experiments with real browsers}
 
With the consideration of real processes of loading a Web page, we conduct our experiments with real browsers for each protocol, respectively. We clone the landing pages of Alexa top 200 websites, which have both mobile version and desktop version, into our local server, and visit them using different protocols through emulated network conditions using \textit{Traffic Control} to keep the consistency across experiments. We vary different packet loss rates, bandwidths, and RTTs to simulate different network conditions and analyze how these network settings affect performance of different protocols. We focus on a user-perceived latency of these pages, so we measure Web page performance using the \textbf{Page Load Time (PLT)} metric. We compare PLTs when loading Web pages using real browsers with HTTP/2, HTTP, and SPDY. PLT is calculated from initiation (when you click on a page link or type in a Web address) to completion (when the page is fully loaded in the browser). We listen the \textit{onLoad} event emitted by browser to get the PLTs.

\subsection{Experiment Settings}

In our experiments, we need the same server implementation to provide HTTP(S), SPDY and HTTP/2 stacks or else our results would be biased. We set up a Nginx proxy~\cite{nginx}, which can transfer Web pages through four protocols, respectively. We need to address that HTTP/2 and SPDY could work without SSL as the standard defines, but both Chrome and Firefox do require the SSL. We employ HTTP/2 and SPDY with encryption. We use Chrome for Android platform to load Web pages we collect, which supports both all the four protocols. We develop a measurement software communicating with Chrome with Chrome Debugging Protocol\footnote{We can communicate with Chrome using Chrome Debugging Protocol to execute commands and listen to events emitted when involving page load, \url{https://developer.chrome.com/devtools/docs/protocol/0.1/index}} to automate our experiments. For each run, we set up network conditions using \textit{Traffic Control} (TC), and control Chrome to load specific Web page. Meanwhile, our measurement software will automatically export a \textit{HTTP archive record} (HAR) format file when page load finished, which record all the requests and responses involved in loading a Web page as well as PLT. 
 
Thus, we need to keep the consistency across experiments. We clone the landing pages of the Alexa top 200 websites that have corresponding mobile version into our local server and convert all external links to local links. Overall, we collect 400 Web pages, consisting of 200 desktop pages and 200 corresponding mobile pages. Meanwhile, we use Linux Traffic Control (TC) to emulate different network to ensure browsers load pages under consistent network conditions for each protocol.

\begin{table}[!htbp]\centering
\caption{Factors may affect HTTP/2 and HTTPS performance.}\label{tab:factors}
\normalsize
\begin{tabular}{|c|c|c|}
\toprule
    \textbf{Factor} & \textbf{Range} & \textbf{High}\\
\midrule
    \textbf{RTT} & 20ms, 100ms, 200ms & $ \geq 100ms $\\
    \hline
    \textbf{bandwidth} & 1Mbps, 10Mbps & $ \geq 10Mbps $\\
    \hline
    \textbf{pkt loss} & 0, 0.005, 0.01, 0.02 & $ \geq 0.01 $\\
    \hline
    \textbf{object size} & 100B, 1K, 10K, 100K, 1M & $ \geq 1K $\\
    \hline
    \textbf{\# of object} & 2, 8, 16, 32, 64, 128, 512 & $ \geq 64 $\\
\bottomrule
\end{tabular}
\end{table}

\textbf{Metrics.} We compare page load time (PLT) when loading Web pages using real browsers with each protocols. PLT is calculated from initiation (when you click on a page link or type in a Web address in the browser) to completion (when the page is fully loaded in the browser). We listen the \textit{onLoad} event emitted by browser to get the PLT. When experiment with our own client, we just download all resources of a page and we calculate the PLT from the time when our client initiates the first request to the time when our client receives the last byte of a page.

\textbf{Server.} We use ThinkPad S3 as our server, which is a 64-bit machine with 1.9GHz 4 core CPU and 8GB memory and Ubuntu 14.04. We could switch the transfer protocol among HTTP, HTTPS, SPDY, and HTTP/2 on Nginx server. We turn off GZIP encoding of Nginx server to keep the exact size of Web objects and Web pages.

\textbf{Client.} We have mentioned we divide our experiments into two parts before. When we compare SPDY, HTTP/2, HTTP, and HTTPS as a transport protocol, we develop our own client based on OkHttp~\cite{okhttp}. For experiment in real browser with different protocols, we load Web pages in Chrome for Android installed in Nexus 6 with 3GB RAM and powered by a 2.7GHz Qualcomm Snapdragon 805 with quad-core CPU (APQ 8084-AB) running Android Lollipop. When analyzing how devices affect the Web page performanceWe, we also conduct experiments using Samsung Galaxy Note 2 with 2 GB RAM and powered by a 1.6GHz quad-core CPU running Android KitKat. By default, we show our experiments conducted on Nexus 6. 

\textbf{Settings.} In this paper, we consider five factors external to HTTP that may affect performance, including network conditions (e.g. packet loss, bandwidth, and RTT), and features of Web pages (e.g. object size, number of objects). Table~\ref{tab:factors} shows the factor settings in our experiments. We use \textit{Traffic Control (TC)} of Linux kernel to set up various network conditions. The RTT values include 20ms (good WiFi), 100ms, and 200ms (3G). The bandwidth (bw) values include 1Mbps (3G) and 10Mbps (WiFi). We vary random packet loss rates from 0\% to 2\%~\cite{Dukkipati:SIGCOMM11}. We also consider a wide range of Web object size (100B-1MB) and number of objects (2-512) when we synthesize Web pages. We define a threshold for each factor, so that we can classify each setting as being high or low for our further analysis.

\textbf{Work flow.} For experiment with real browsers, we develop a measurement software communicating with Chrome with Chrome Debugging Protocol to automate our experiments. For each run, we set up network conditions using TC, and control Chrome to load specific Web page. For example, we send ``Page.navigation'' command with specific URL to chrome, and Chrome will load the target Web page. Meanwhile, our measurement software will automatically export a HAR-format file when page load finished, which record all the requests and responses involved in loading a Web page as well as PLT. We will clear browsers' cache before loading another Web page. For experiments with our own client, our client will automatically download all pages and record the PLT for each page.

%% file: section/performance_protocols.tex
\section{Performance as Transfer Protocols }
\begin{figure}[htbp]
    \includegraphics[width=1.0\textwidth]{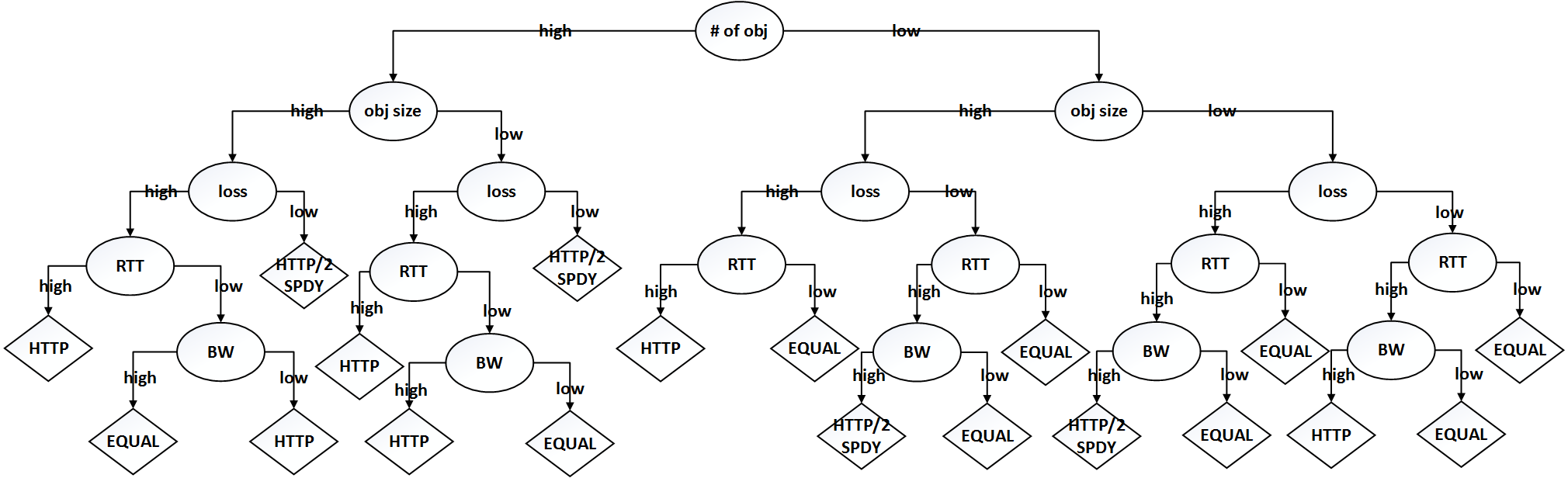}
    \caption{The decision tree that tells how HTTP/2, SPDY, and HTTP perform over parameter space.} \label{fig:decision_tree}
\end{figure}

In this section, we compare the performance when consider four protocols as transfer protocols to access Web contents hosted in remote servers. We need to eliminate the effect of dependencies and computation in real browsers, so we develop our own mobile client to load pages, which just downloads all resources of a page and records the timeline of resources downloading. We consider a wide range of parameter settings as showed in Table~\ref{tab:factors}. Our client will fetch all the resources from these synthetic pages with pre-specified object size and object number. We switch transfer protocol of Nginx proxy among HTTP, HTTPS, SPDY, and HTTP/2, so that our mobile client can work with these protocols, respectively. 

To understand how these four protocols perform under different conditions, we try to build a predictive model based on decision tree analysis. We have mentioned that HTTPS piggybacks HTTP entirely on top of TLS/SSL, which need one more expensive TLS/SSL handshake for each new connection with extra two Round-Trip Time (RTT). HTTPS do have worse performance that HTTP, and we just conduct the decision tree analysis among HTTP, SPDY, and HTTP/2.

In the analysis, each configuration is a combination of values for all factors listed in Table~\ref{tab:factors}. For each configuration, we add an additional variable \textit{s}, which means the performance of these protocols. We consider a protocol better than another if at least in the 70\% of the measurement cases performed at least 10\% faster than the other protocol’s average PLT. If two of the protocols are fulfilling this condition against the third one but not against each other we marked both of them. In case of this condition doesn’t stand between any of the three protocols we marked them as equal. We run the decision tree with ID3 algorithm~\cite{decisiontree} to predict the factor settings under which HTTP/2 performs better (or HTTP/SPDY performs better). The ID3 algorithm begins with the original set \textit{S} as the root node. On each iteration of the algorithm, it iterates through every unused factor (including RTT, bw, packet loss rate, object size, and object number in our data set) of the set \textit{S} and calculates information gain \textit{IG(A)} of that attribute. It then selects the attribute which has the largest information gain value. The set \textit{S} is then split by the selected attribute to produce subsets of the data. The algorithm continues to recurse on each subset, considering only attributes never selected before. 

Figure~\ref{fig:decision_tree} shows the decision tree that tells how HTTP/2, SPDY, and HTTP perform over parameter space. We could see that SPDY and HTTP/2 have similar performance across different settings. It is not surprising since HTTP/2 derived from SPDY and they share similar implementation. We have some conclusions as following:

\begin{itemize}
	\item{HTTP/2 and SPDY do not always gain performance improvement under different parameter settings. }
    \item{HTTP/2 and SPDY perform worse with high packet loss. HTTP/2 and SPDY both have the feature of one TCP connection per origin. A single connection hurts under high packet loss because it aggressively reduces the congestion window compared to HTTPS which reduces the congestion window on only one of its parallel connections. When this single connection suffers from packet loss, all streams running over this unique TCP connection are negatively impacted.}
    \item{HTTP/2 and SPDY perform better with many objects under low loss. TCP implements congestion control by counting outstanding packets not bytes. Thus, sending a few small objects with HTTP will promptly use up the congestion window, though outstanding bytes are far below the window limit. In contrast, A single HTTP/2 connection can contain multiple concurrently-open streams, with either endpoint interleaving frames from multiple streams.}
    \item{High RTT favors HTTP/2 and SPDY against HTTP due to multiplexing. HTTP/2 and SPDY benefits from having a single connection and stream multiplexing. One connection per origin significantly reduces the associated overhead: fewer sockets to manage along the connection path, smaller memory footprint, better connection throughput, less time in slow-start, faster congestion and loss recovery. As the RTT goes up, new established TCP connections cost more time.}
\end{itemize}

The decision tree also depicts the relative importance of contributing factors. Intuitively, factors close to the root of the decision tree affect HTTP/2‘ performance more than those near the leaves. This is because the decision tree places the important factors near the root to reduce the number of branches. We find that object number, object size, and packet loss rate are the most important factors in predicting HTTP/2‘ performance. However, RTT and bandwidth play a less important role as shown in Figure~\ref{fig:decision_tree}.

%% file: section/web_page_performance.tex
\section{Web page performance under multiple protocols}

In this section, we analyze the impact of multiple protocols under different network conditions on Web page performance when loading pages in real browsers. We also conduct our experiments on Samsung Galaxy Note 2 to analyze the impact of different devices on Web page performance. We find that Web page performances differ marginally between these two devices.

\subsection{Characterizing Web pages}
\begin{figure}[htbp]
\centering
\subfigure[\# Object]{
\label{fig:object_num}
\includegraphics[width=0.32\textwidth]{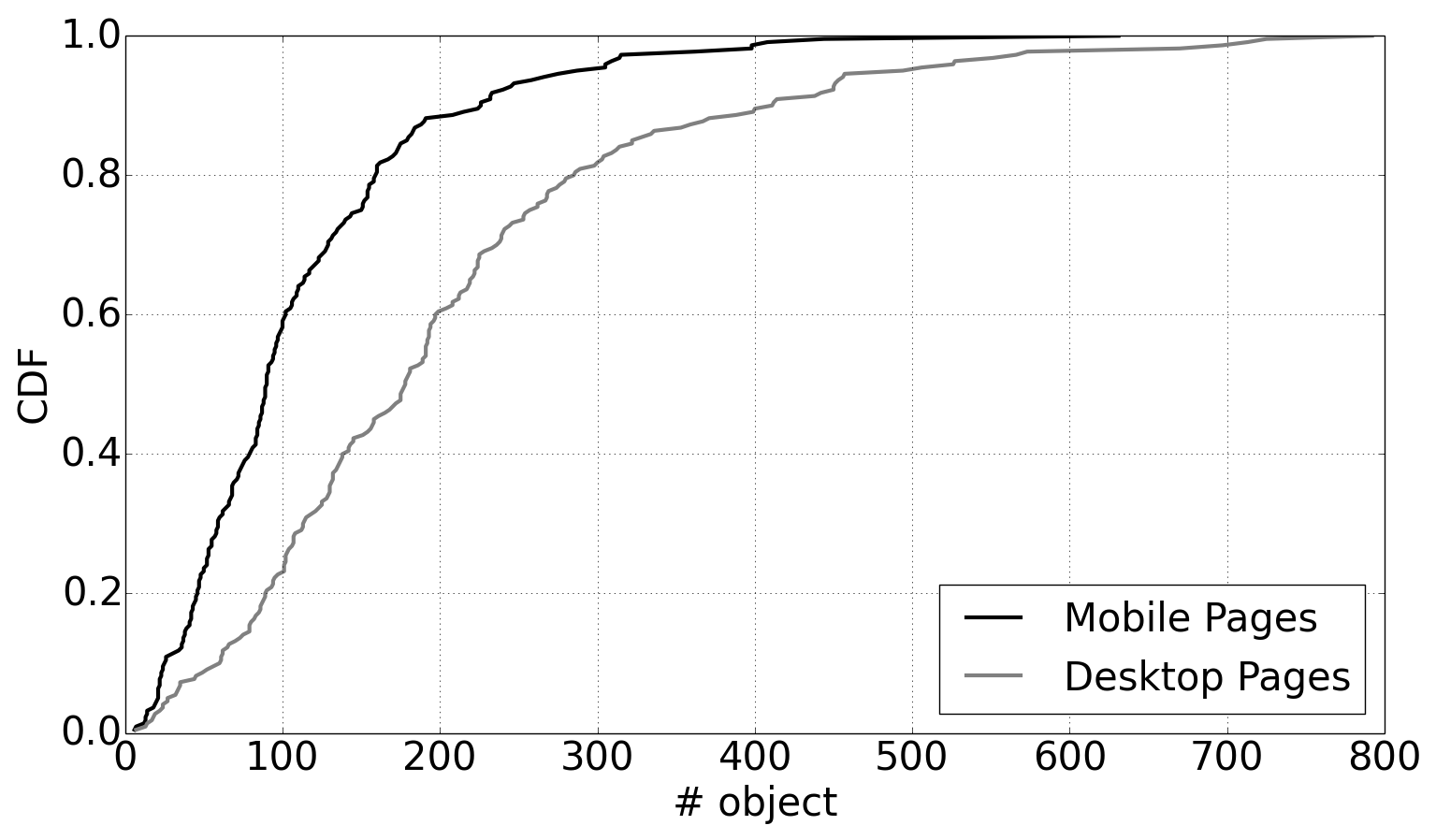}}
\subfigure[Page Size]{
\label{fig:page_size}
\includegraphics[width=0.32\textwidth]{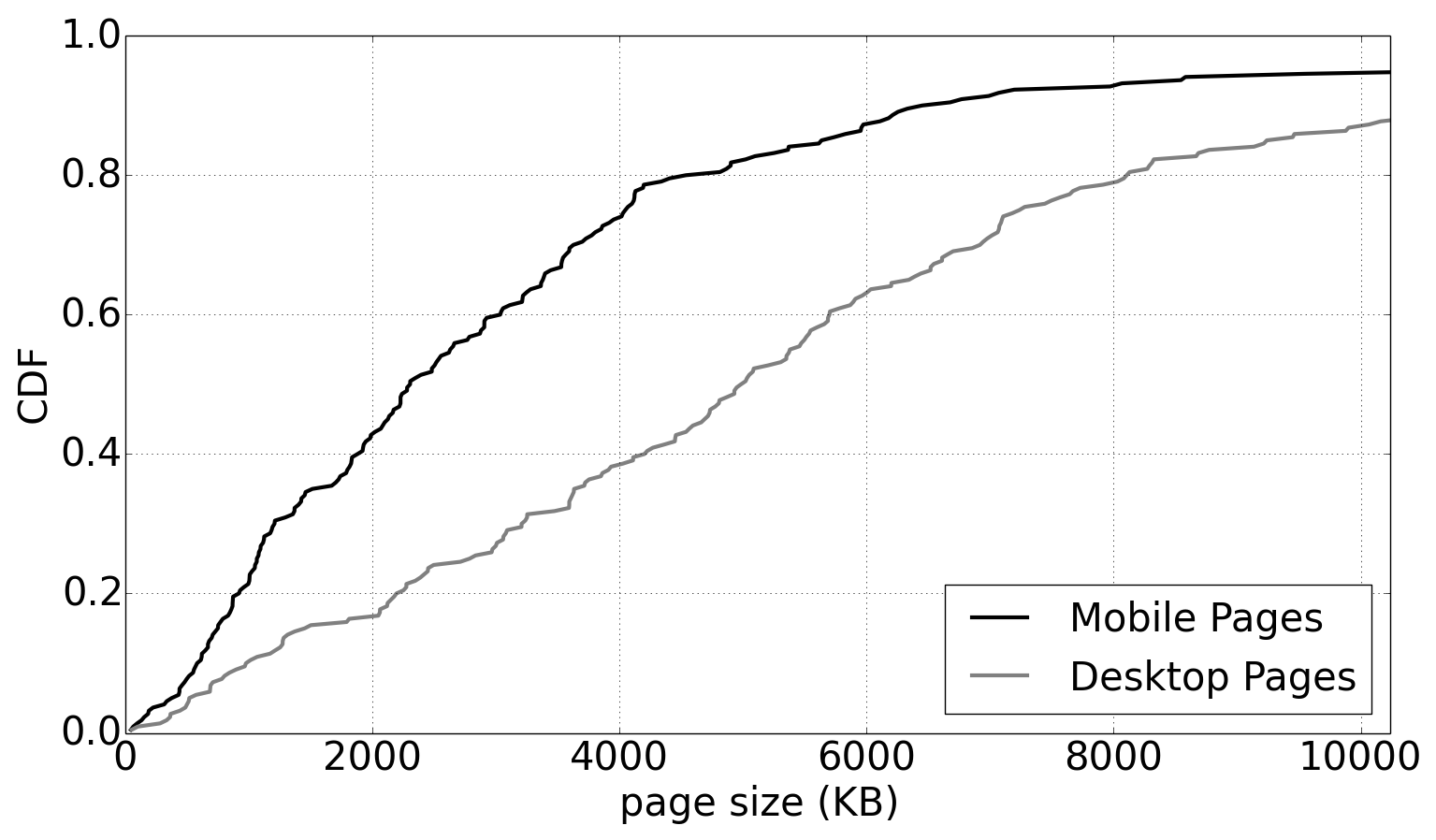}}
\subfigure[Overall Object Size]{
\label{fig:overall_object_size}
\includegraphics[width=0.32\textwidth]{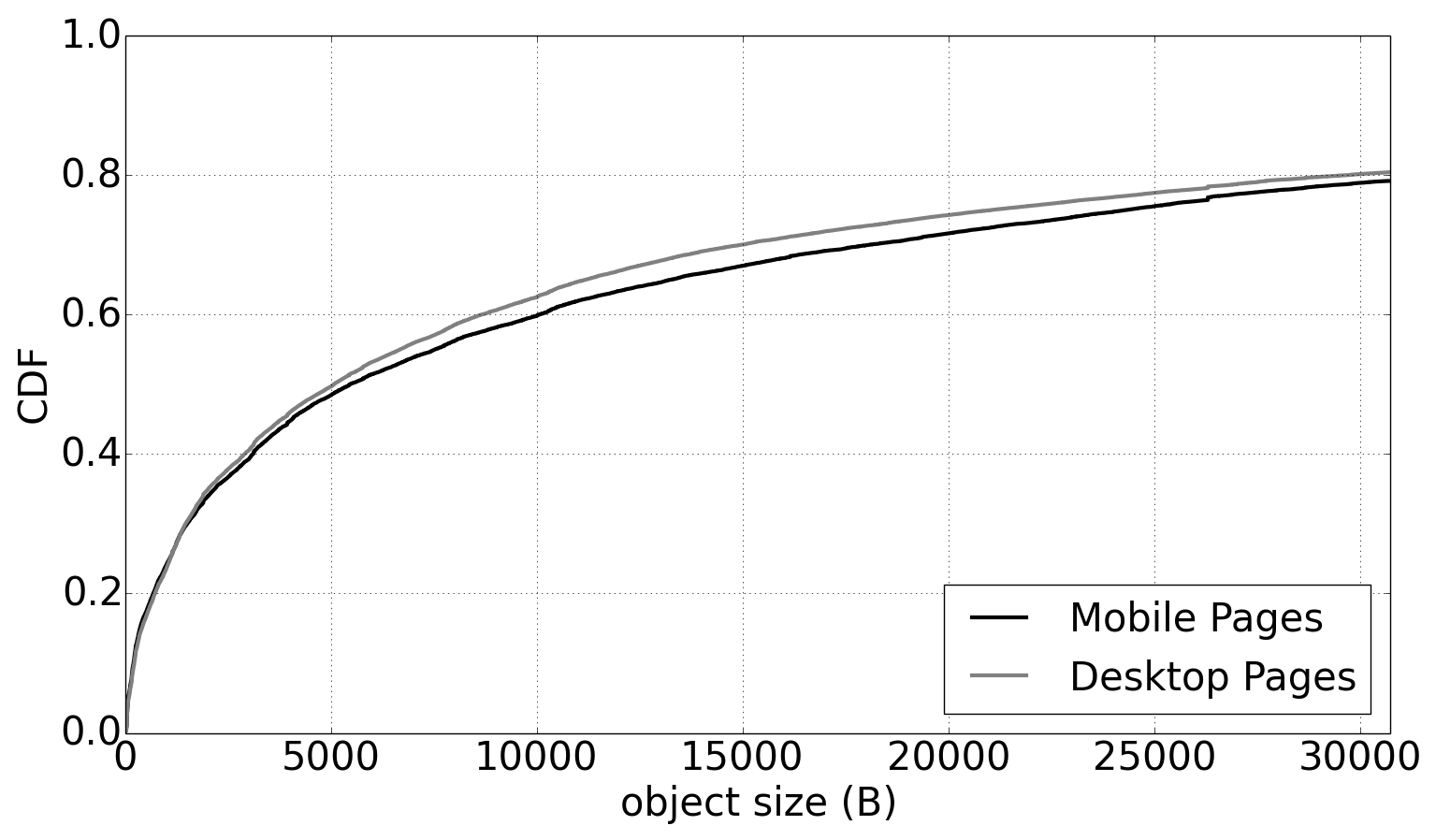}}
\caption{Characteristics of Web pages, including desktop pages and mobile pages}
\label{fig:characteristics_of_pages}
\end{figure}

First, we analyze the complexity of Web pages we collect. Especially, we focus on the differences between desktop pages and mobile pages. Original Web pages are developed for desktop browsers, and contains large number of objects to provide abundant contents. With the popularity of mobile devices, users are used to visiting Web pages using mobile browsers. However, mobile devices have smaller screen and limited resources including computing capacity, power supply, and data plan. It may not be suitable to present desktop pages in mobile browser. Many Web developers maintain a mobile version of pages to provide better user experience. 

We collect landing pages of Alexa Top 200 websites that have corresponding mobile version. We need address that we do not collect those pages with responsive Web design. A responsive Web page may present differently on mobile browser, but it does download a same collection of resources as desktop browser. Overall, we have 400 Web pages, consisting of 200 desktop pages and 200 corresponding mobile pages. 

Figure~\ref{fig:characteristics_of_pages} shows the cumulative distribution function (CDF) of  characteristics of Web pages including desktop version and mobile version. We could see that mobile pages does reduce the number of objects and overall page size in Figure~\ref{fig:object_num} and Figure~\ref{fig:page_size}. From Figure~\ref{fig:object_num}, We can see that the median object number of mobile pages is 90, but the median object number of desktop page is 178. Meanwhile, the median page size of mobile pages is 2.25MB, but the median page size of desktop page is 4.88MB. We can see that mobile pages are optimized with less objects and smaller page size and reduce nearly half of objects and page sizes compared to desktop pages in the median case. Figure~\ref{fig:overall_object_size} shows the distribution of object size across all Web pages we collect for desktop pages and mobile pages, respectively. We can find that the distributions of object size for desktop pages and mobile pages are similar.

\begin{figure}[htbp]
	\centering
    \includegraphics[width=0.6\textwidth]{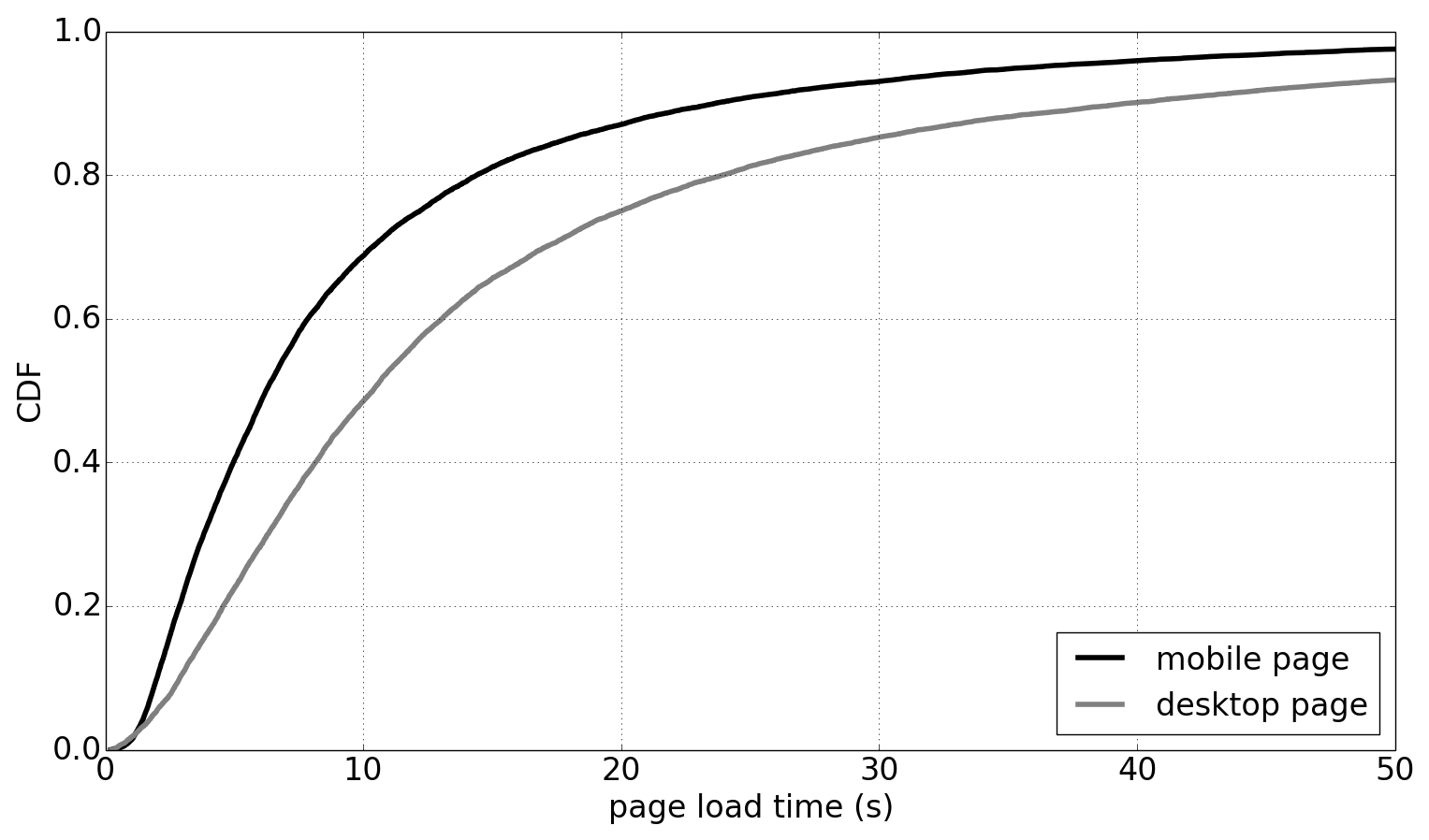}
    \caption{Page load times for desktop pages and mobile pages. Results are across network conditions with four protocols.} \label{fig:pagetype_pageload_overall}
\end{figure}

Figure~\ref{fig:pagetype_pageload_overall} shows the page load times on mobile browser for desktop pages and mobile pages, respectively. The results are across all network conditions with four protocols as we present in last section. In the median case, page load times are 6.08s for mobile pages and 10.06s for desktop pages, and browsers could save about 40\% of time when loading mobile pages.

\subsection{Web page performance with different protocols}

\begin{figure*}[tbp]
\centering
\subfigure[PLT with HTTP under different RTTs]{
\label{fig:rtt_http}
\includegraphics[width=0.32\textwidth]{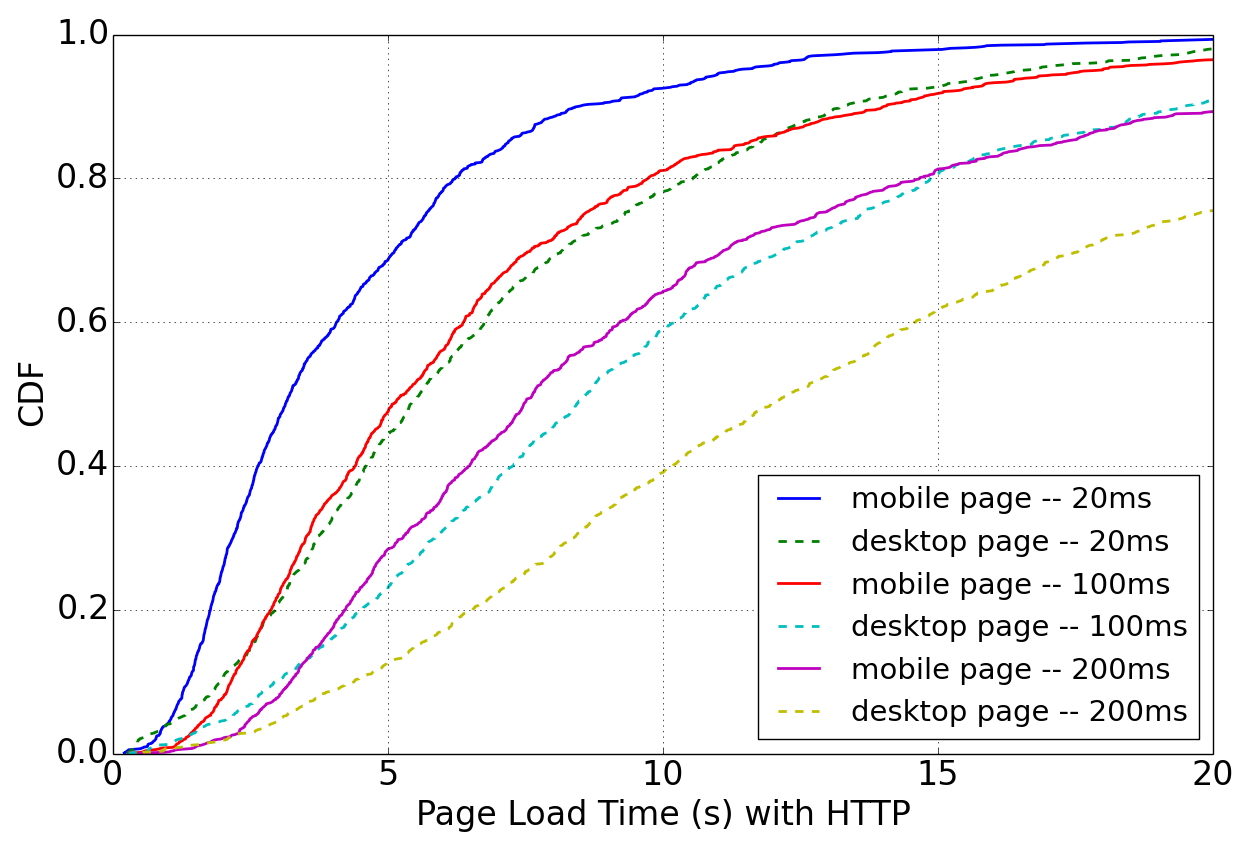}}
\subfigure[PLT with HTTP under different bandwidths]{
\label{fig:bw_http}
\includegraphics[width=0.32\textwidth]{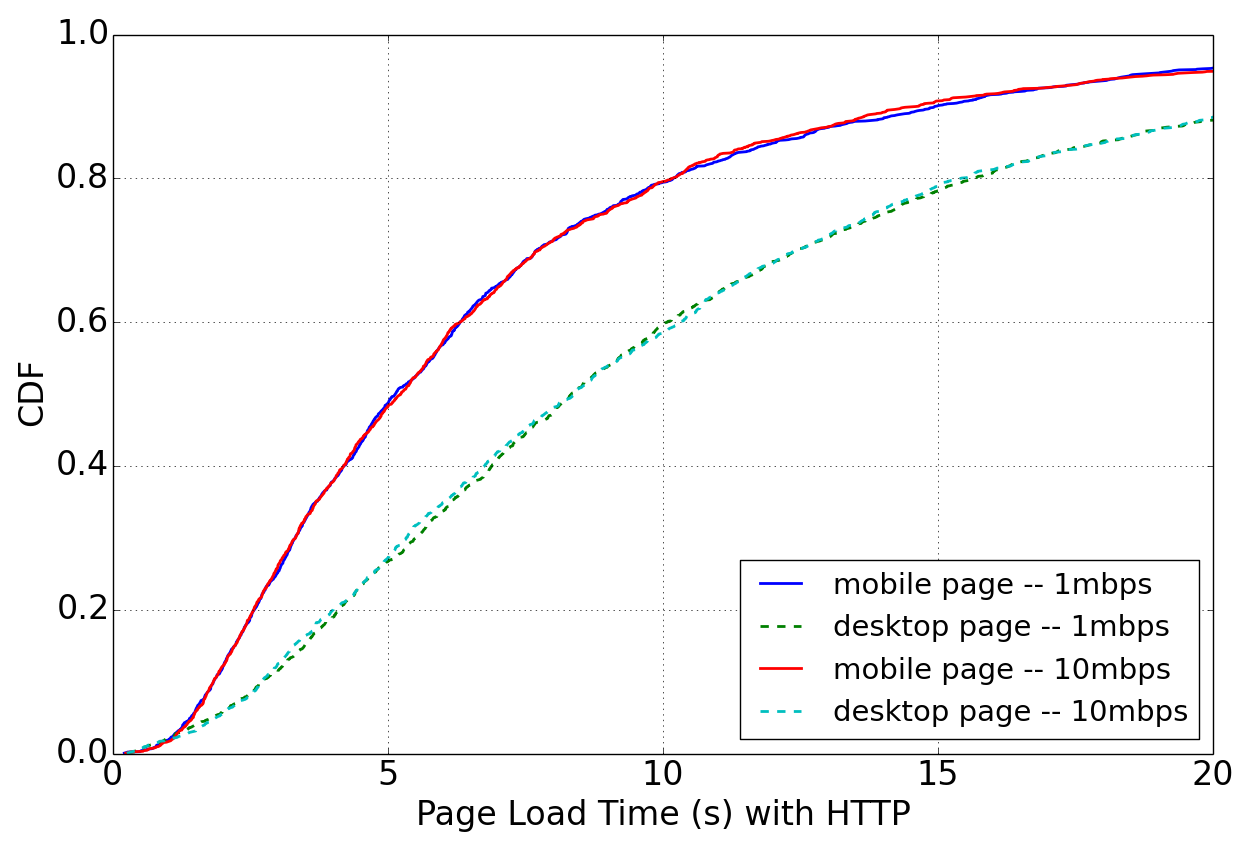}}
\subfigure[PLT with HTTP under different packet loss rate]{
\label{fig:loss_http}
\includegraphics[width=0.32\textwidth]{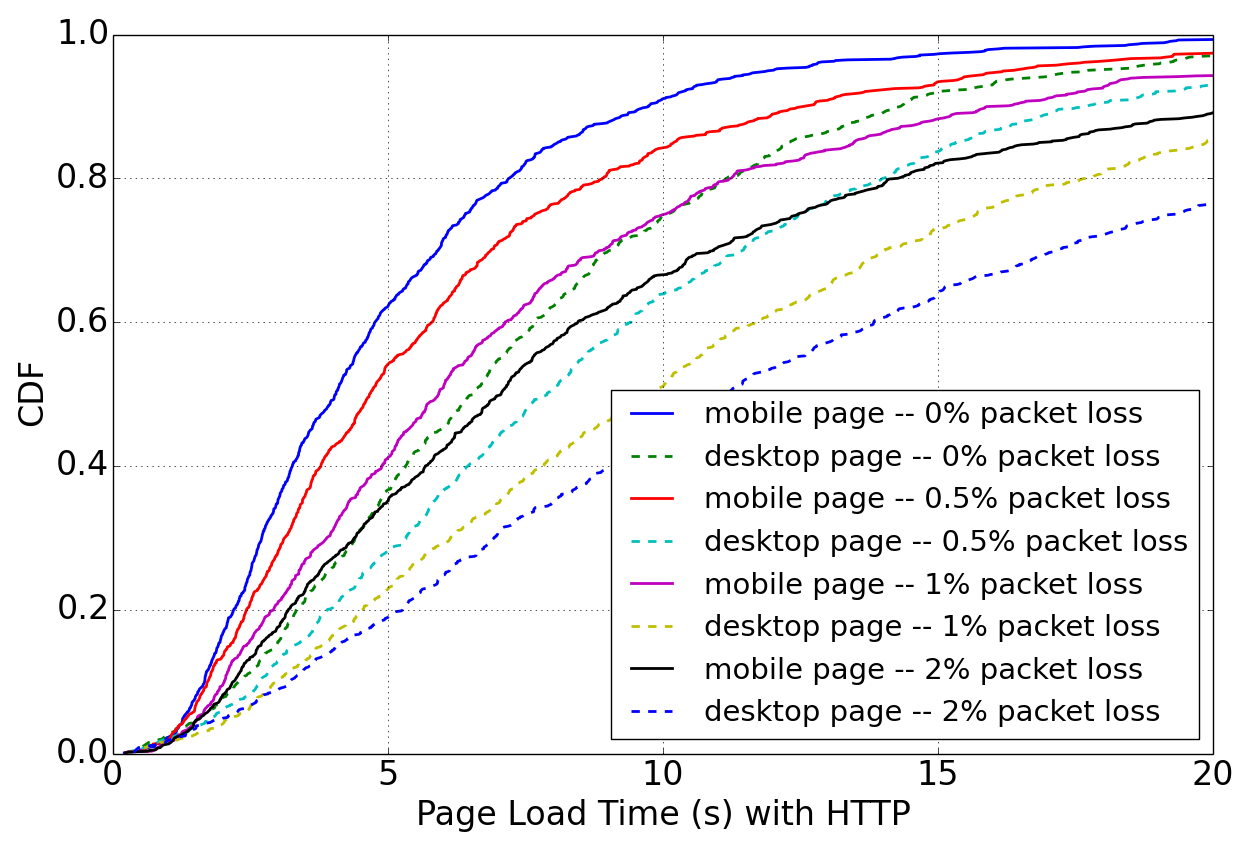}}
\caption{PLT with HTTP under different network .}
\label{fig:plt_with_http}
\end{figure*}

\begin{figure*}[tbp]
\centering
\subfigure[PLT with HTTPS under different RTTs]{
\label{fig:rtt_https}
\includegraphics[width=0.32\textwidth]{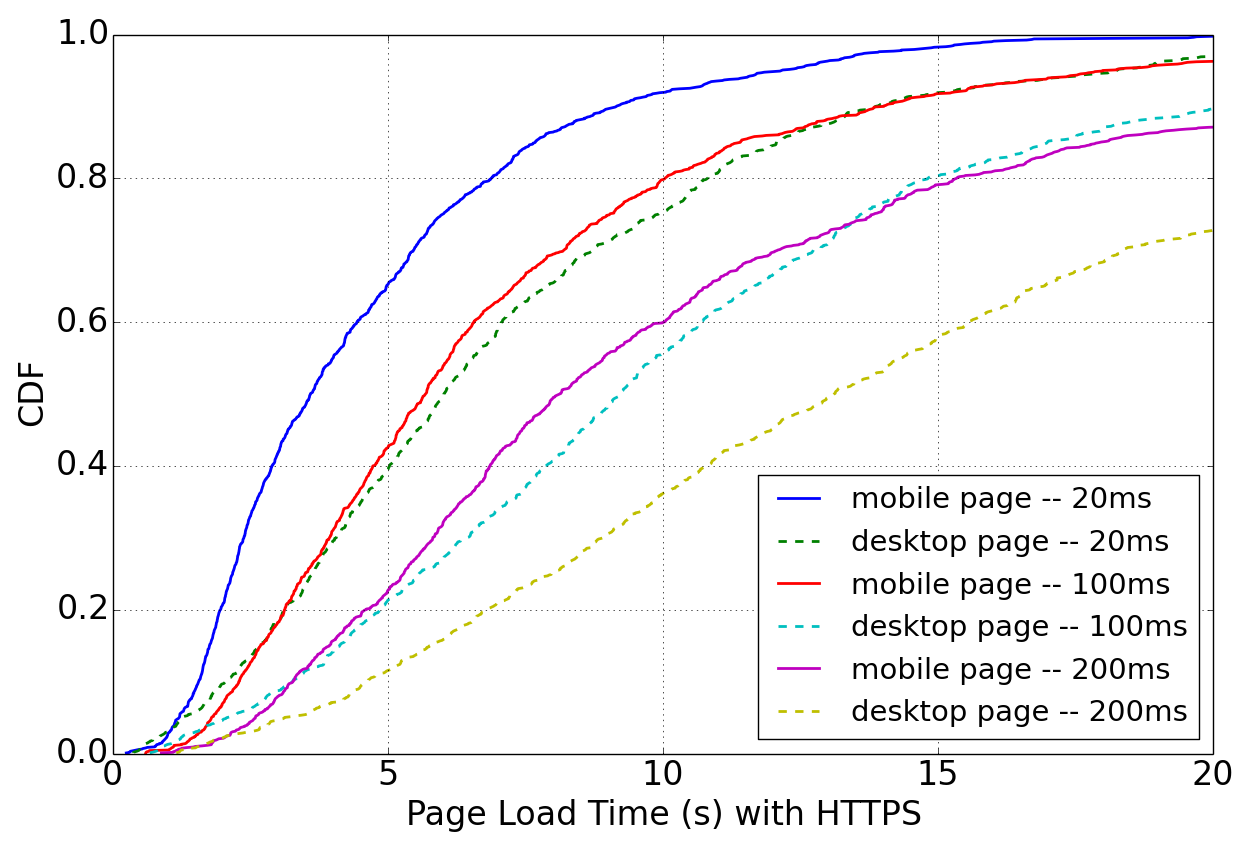}}
\subfigure[PLT with HTTPS under different bandwidths]{
\label{fig:bw_https}
\includegraphics[width=0.32\textwidth]{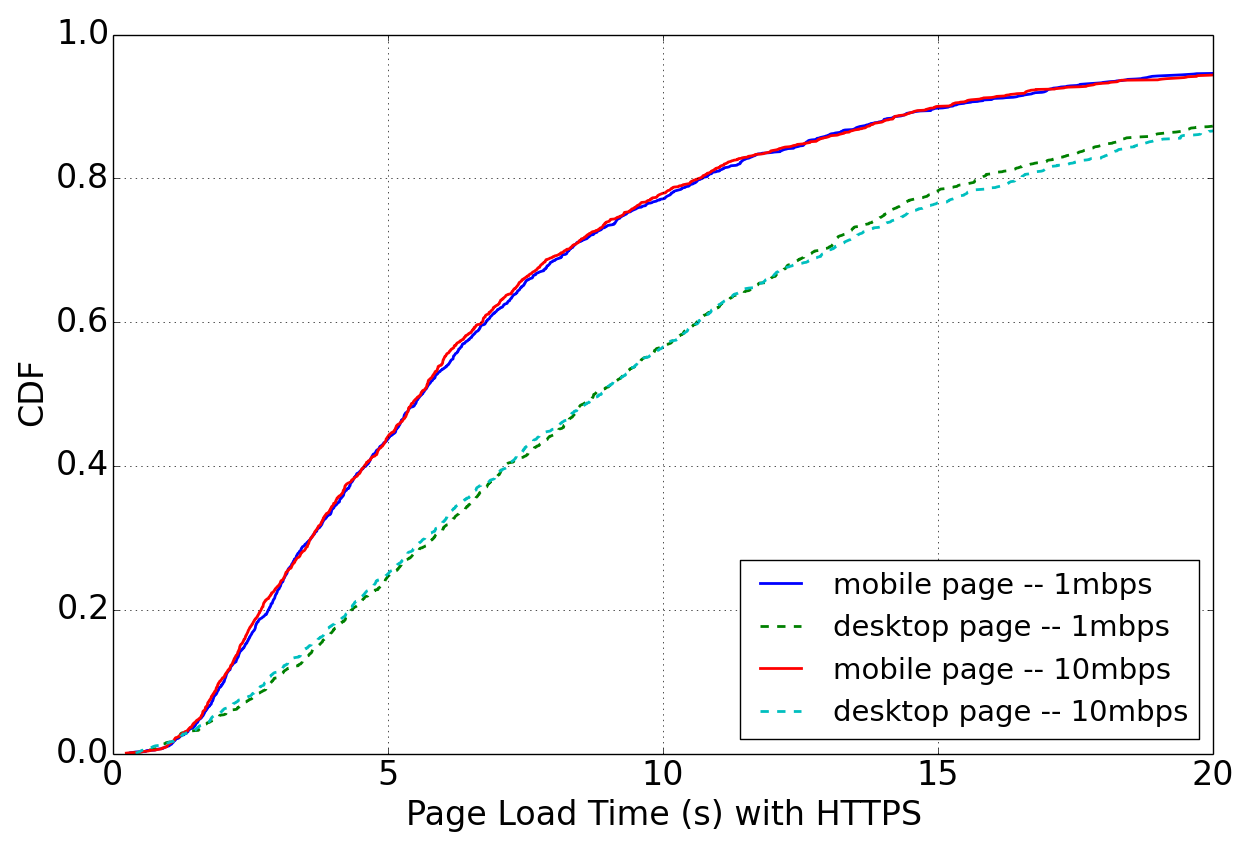}}
\subfigure[PLT with HTTPS under different packet loss rate]{
\label{fig:loss_https}
\includegraphics[width=0.32\textwidth]{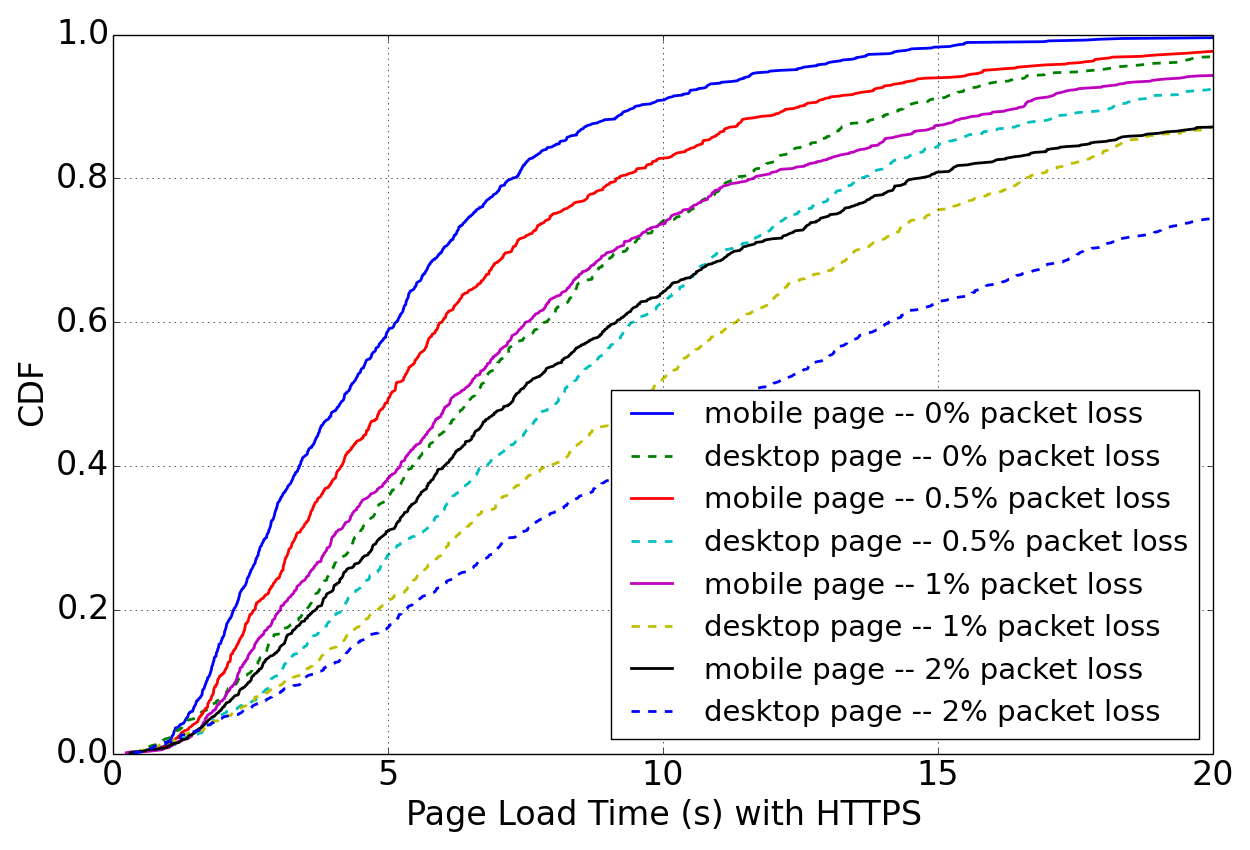}}
\caption{PLT with HTTPS under different network .}
\label{fig:plt_with_https}
\end{figure*}

\begin{figure}[htbp]
\centering
\subfigure[PLT with HTTP/2 under different RTTs]{
\label{fig:rtt_http2}
\includegraphics[width=0.32\textwidth]{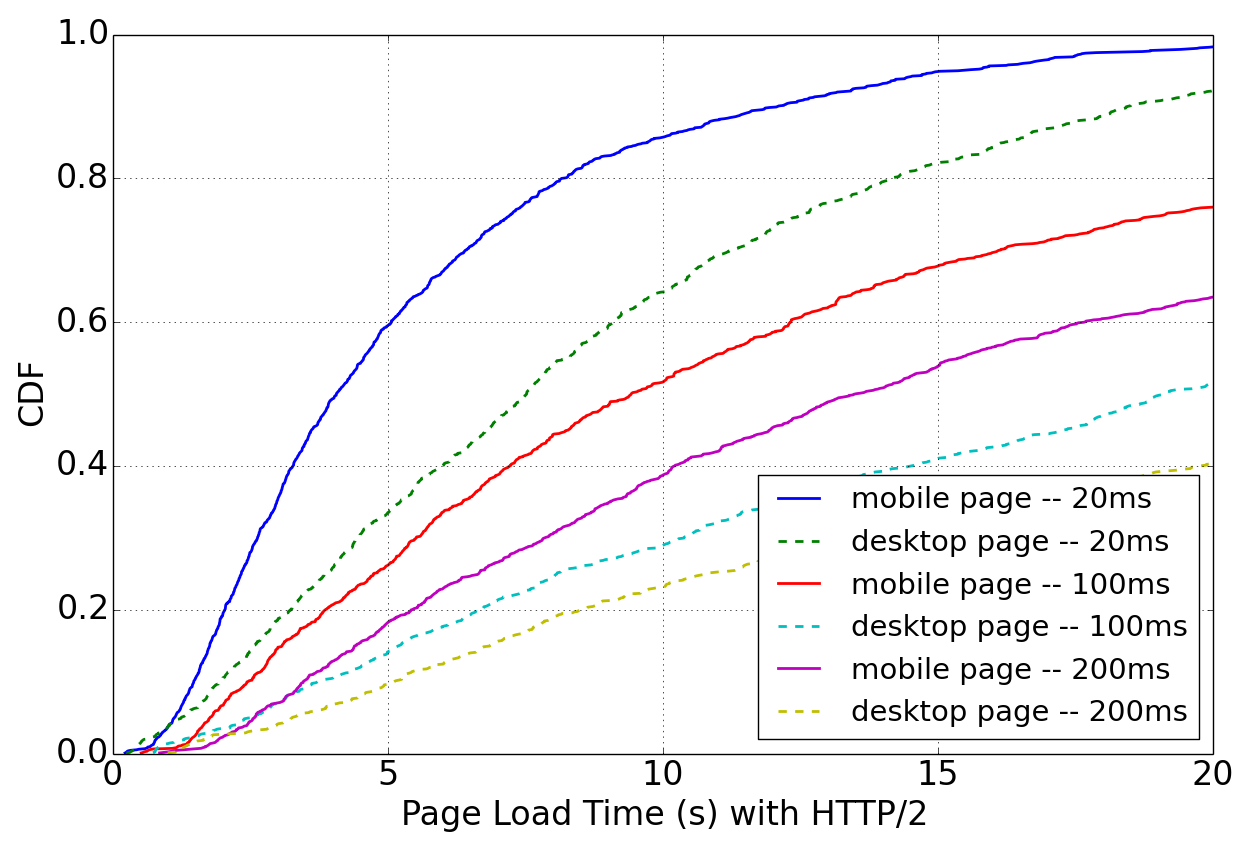}}
\subfigure[PLT with HTTP/2 under different bandwidths]{
\label{fig:bw_http2}
\includegraphics[width=0.32\textwidth]{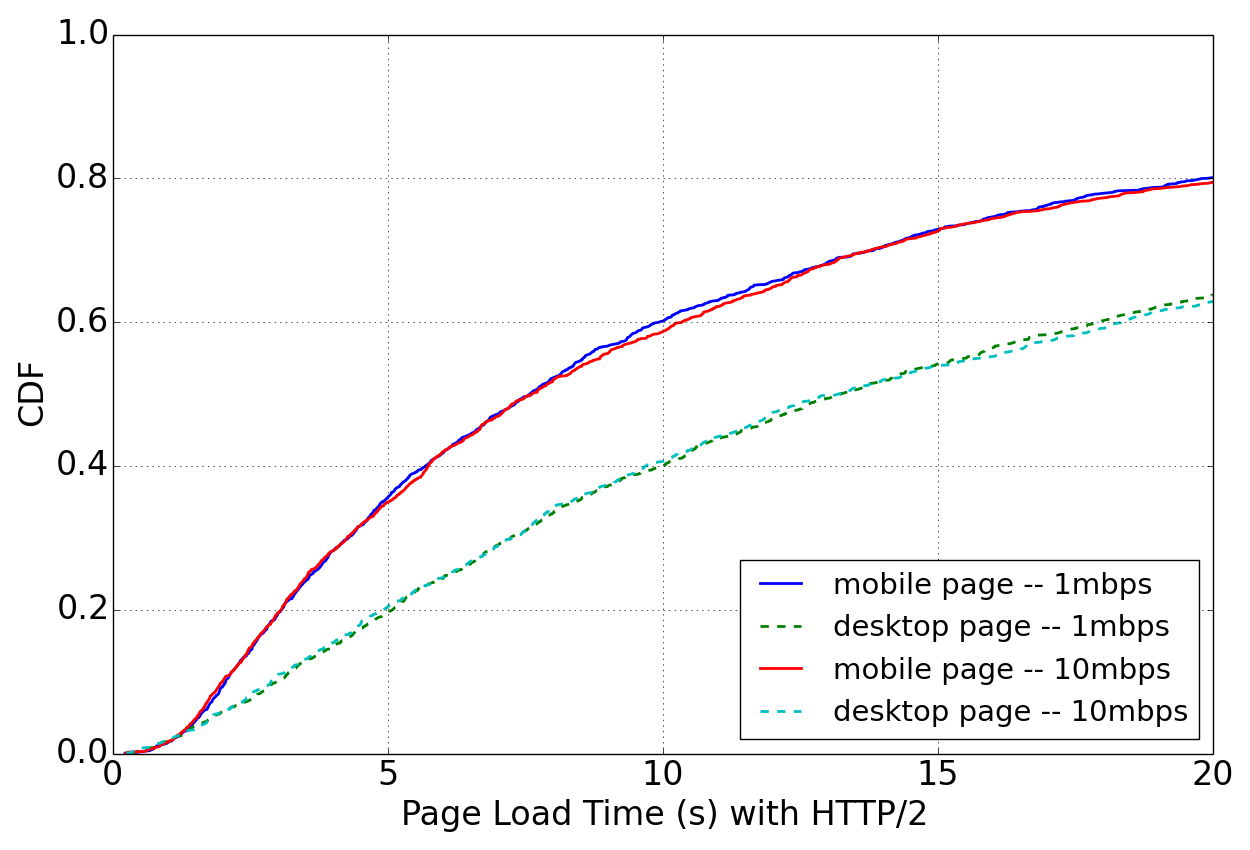}}
\subfigure[PLT with HTTP/2 under different packet loss rate]{
\label{fig:loss_http2}
\includegraphics[width=0.32\textwidth]{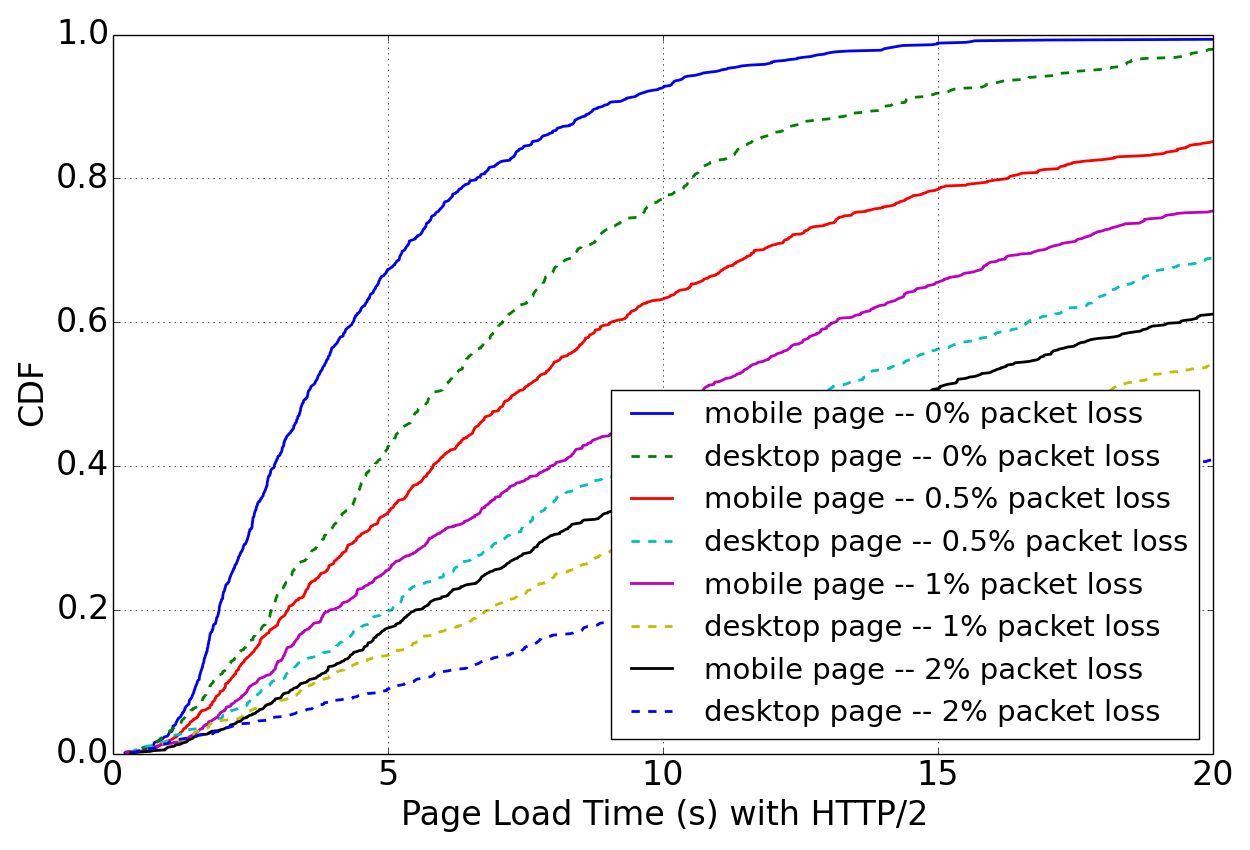}}
\caption{PLT with HTTP/2 under different network .}
\label{fig:plt_with_http2}
\end{figure}

\begin{figure}[htbp]
\centering
\subfigure[PLT with SPDY under different RTTs]{
\label{fig:rtt_spdy}
\includegraphics[width=0.32\textwidth]{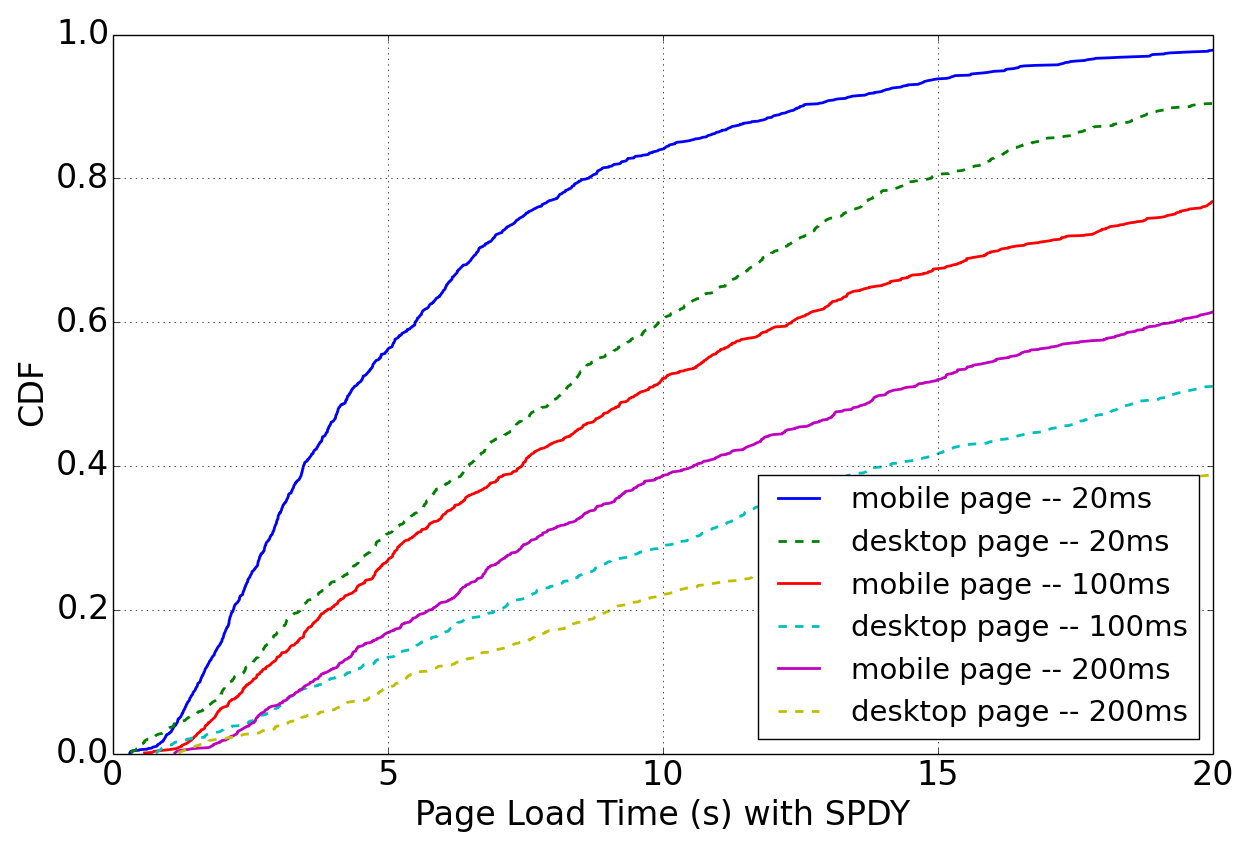}}
\subfigure[PLT with SPDY under different bandwidths]{
\label{fig:bw_spdy}
\includegraphics[width=0.32\textwidth]{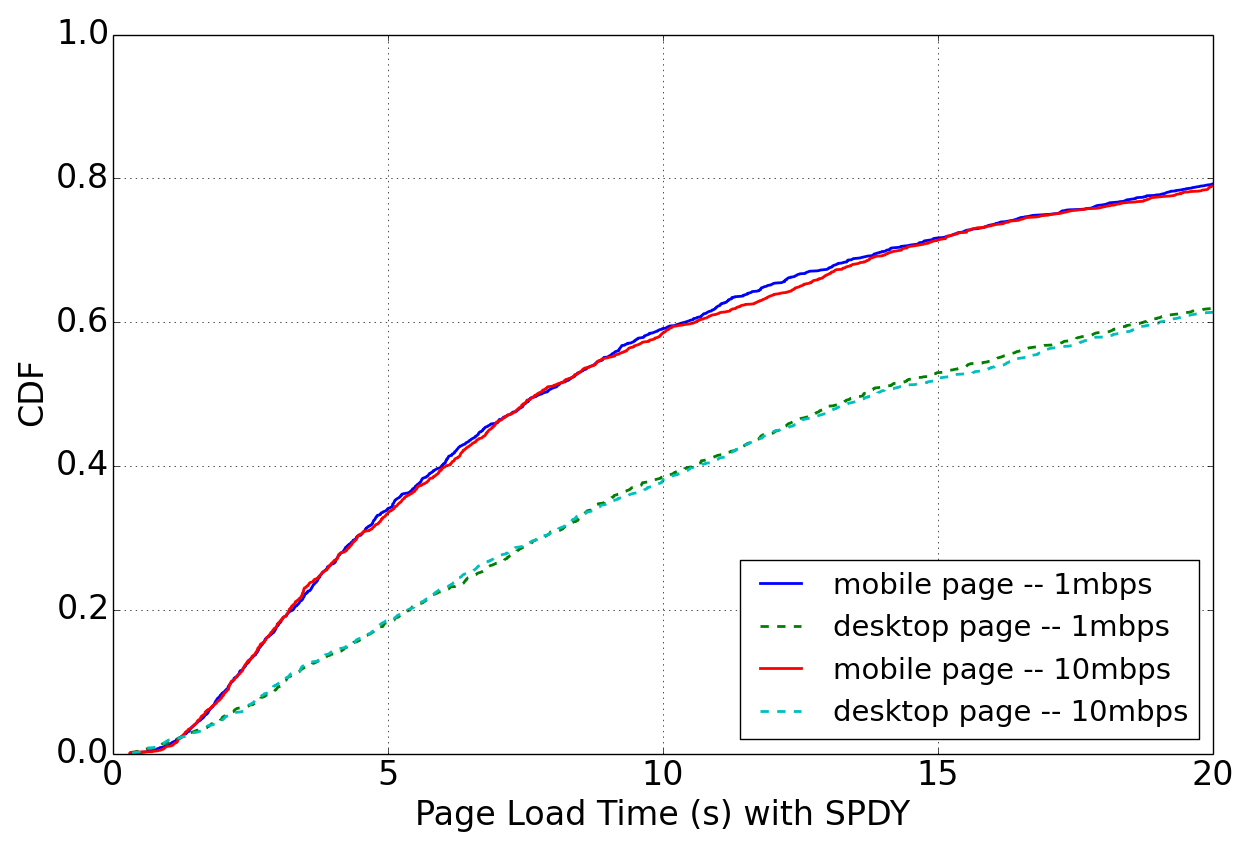}}
\subfigure[PLT with SPDY under different packet loss rate]{
\label{fig:loss_spdy}
\includegraphics[width=0.32\textwidth]{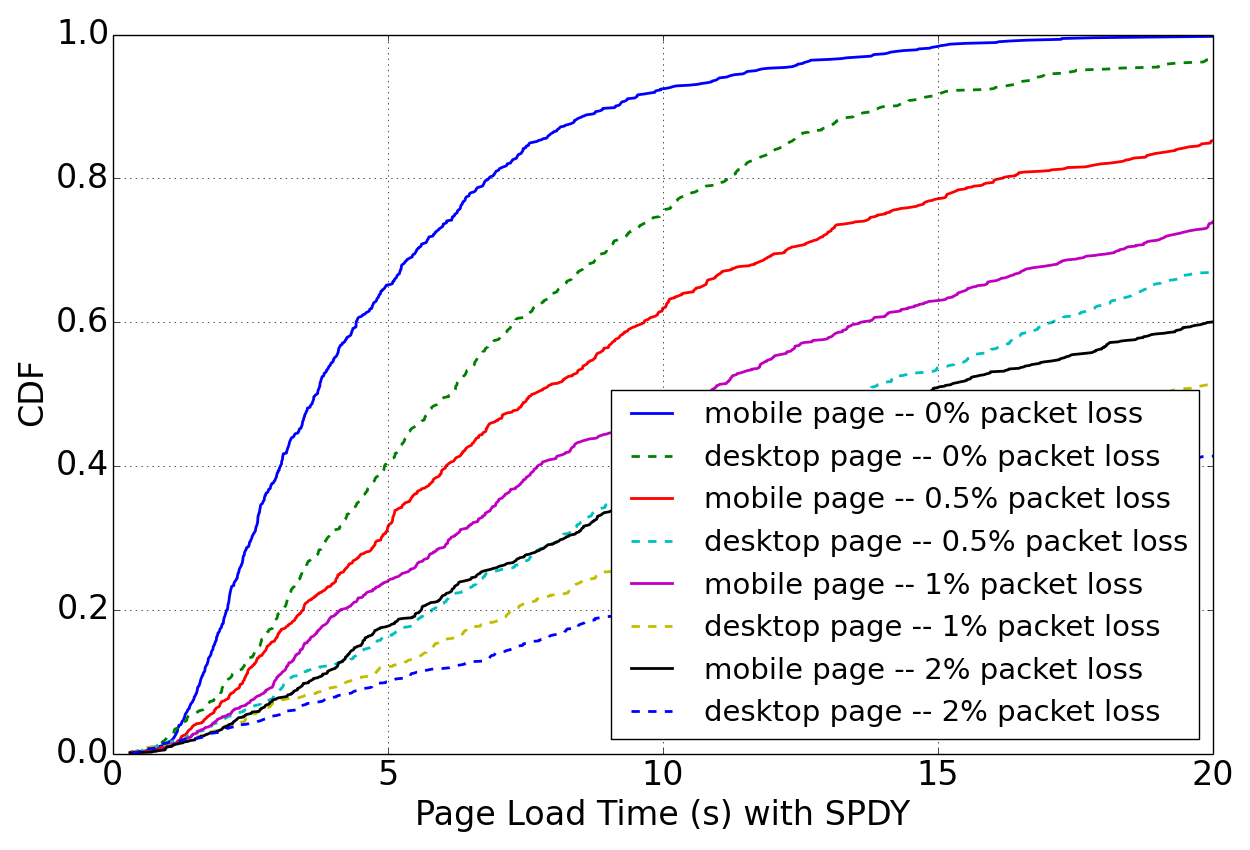}}
\caption{PLT with SPDY under different network .}
\label{fig:plt_with_spdy}
\end{figure}

In this section, we analyze the impact of four protocols under different network conditions on Web page performance when loading pages in real browsers. We focus on how Web page performance may be affected for each protocol under different network conditions as shown in Figure~\ref{fig:plt_with_http}, Figure~\ref{fig:plt_with_https}, Figure~\ref{fig:plt_with_http2}, and Figure~\ref{fig:plt_with_spdy}, respectively.

Figure~\ref{fig:plt_with_http} shows the distribution of PLT with HTTP under different network conditions. Figure~\ref{fig:rtt_http}, Figure~\ref{fig:bw_http}, and Figure~\ref{fig:loss_http} show how PLTs vary under different RTTs, different bandwidths, and different packet loss rates for both mobile pages and desktop pages, respectively. In the median case, mobile pages can reduce 37.3\% $\sim$ 42.3\% PLT compared to desktop pages as RTTs arise, and mobile pages perform better with higher RTT. Figure~\ref{fig:bw_http} shows that bandwidths do not affect PLT for both mobile pages and desktop pages. Figure~\ref{fig:loss_http} shows that PLTs increase as packet loss rates arise. Performance of mobile pages with 2\% packet loss rate could even approach performance of desktop pages with 0\% packet loss rate. We recommend adaption of mobile pages for developers to provide better user experience even under bad network conditions.

Figure~\ref{fig:plt_with_https} shows the distribution of PLT with HTTPS under different network conditions. Figure~\ref{fig:rtt_https}, Figure~\ref{fig:bw_https}, and Figure~\ref{fig:loss_https} show how PLTs vary under different RTTs, different bandwidths, and different packet loss rates for both mobile pages and desktop pages, respectively. In the median case, Web pages load with HTTPS cost extra 6.5\% $\sim$ 11.5\% time since HTTPS costs extra two RTTs for establishing a new connection. Overall, PLTs with HTTPS have a similar trends as PLTs with HTTP. Thus, Bandwidths do not have significant affect on PLTs with HTTPS, too. Figure~\ref{fig:rtt_https} shows that in the median case, mobile pages can reduce 38.0\% $\sim$ 40.3\% PLT compared to desktop pages as RTTs arise.

Figure~\ref{fig:plt_with_http2} shows the distribution of PLT with HTTP/2 under different network conditions. Figure~\ref{fig:rtt_http2}, Figure~\ref{fig:bw_http2}, and Figure~\ref{fig:loss_http2} show how PLTs vary under different RTTs, different bandwidths, and different packet loss rates for both mobile pages and desktop pages, respectively. Bandwidths do not have significant affect on PLTs with HTTP/2, too. However, we find that PLTs with HTTP/2 are more sensitive to other network parameters. As the RTT changes from 20ms to 100ms, PLTs double for both desktop pages and mobile pages. In the median case, mobile pages can reduce 45.9\% $\sim$ 50.6\% PLT compared to desktop pages as RTTs arise. In the median case, HTTP/2 reduces 12.5\% and 11.2\% PLTs against HTTP for mobile pages and desktop pages, respectively when packet loss rate is 0\%. However, HTTP/2 performs worse as packet loss rate arise. When packet loss rate arises to 2\%, PLT of HTTP/2 is almost twice as long as that of HTTP for both mobile pages and desktop pages. As we discuss in previous section, HTTP/2 and SPDY work worse when packet loss rate is high. HTTP/2 and SPDY have the feature of one TCP connection per origin. A single connection hurts under high packet loss because it aggressively reduces the congestion window compared to HTTPS which reduces the congestion window on only one of its parallel connections. When this single connection suffers from packet loss, all streams running over this unique TCP connection are negatively impacted.

Figure~\ref{fig:plt_with_spdy} shows the distribution of PLT with SPDY under different network conditions. Figure~\ref{fig:rtt_spdy}, Figure~\ref{fig:bw_spdy}, and Figure~\ref{fig:loss_spdy} show how PLTs vary under different RTTs, different bandwidths, and different packet loss rates for both mobile pages and desktop pages, respectively. In the median case, mobile pages can reduce 47.2\% $\sim$ 52.1\% PLT compared to desktop pages as RTTs arise. We find that SPDY has a similar performance trend as HTTP/2. We infer that HTTP/2 derived from SPDY and they share a similar implementation in the Nginx server on which we conduct our experiments.

\subsection{Web page performance with different devices}

We conduct our previous controlled experiments on Samsung Galaxy Note 2 to observer if different devices could affect Web page performance. The Nexus 6 phone is equipped with 3GB RAM and powered by a 2.7GHz Qualcomm Snapdragon 805 with quad-core CPU (APQ 8084-AB) running Android Lollipop. The Samsung Galaxy Note 2 phone is equipped with 2 GB RAM and powered by a 1.6GHz quad-core CPU running Android KitKat. We can see that Samsung Galaxy Note 2 has fewer RAM and weaker CPU.

\begin{figure}[htbp]
\centering
\subfigure[Page load time when loading pages in Nexus 6 phones and Galaxy Note 2 phones with HTTP]{
\label{fig:device_pageload_http}
\includegraphics[width=0.48\textwidth]{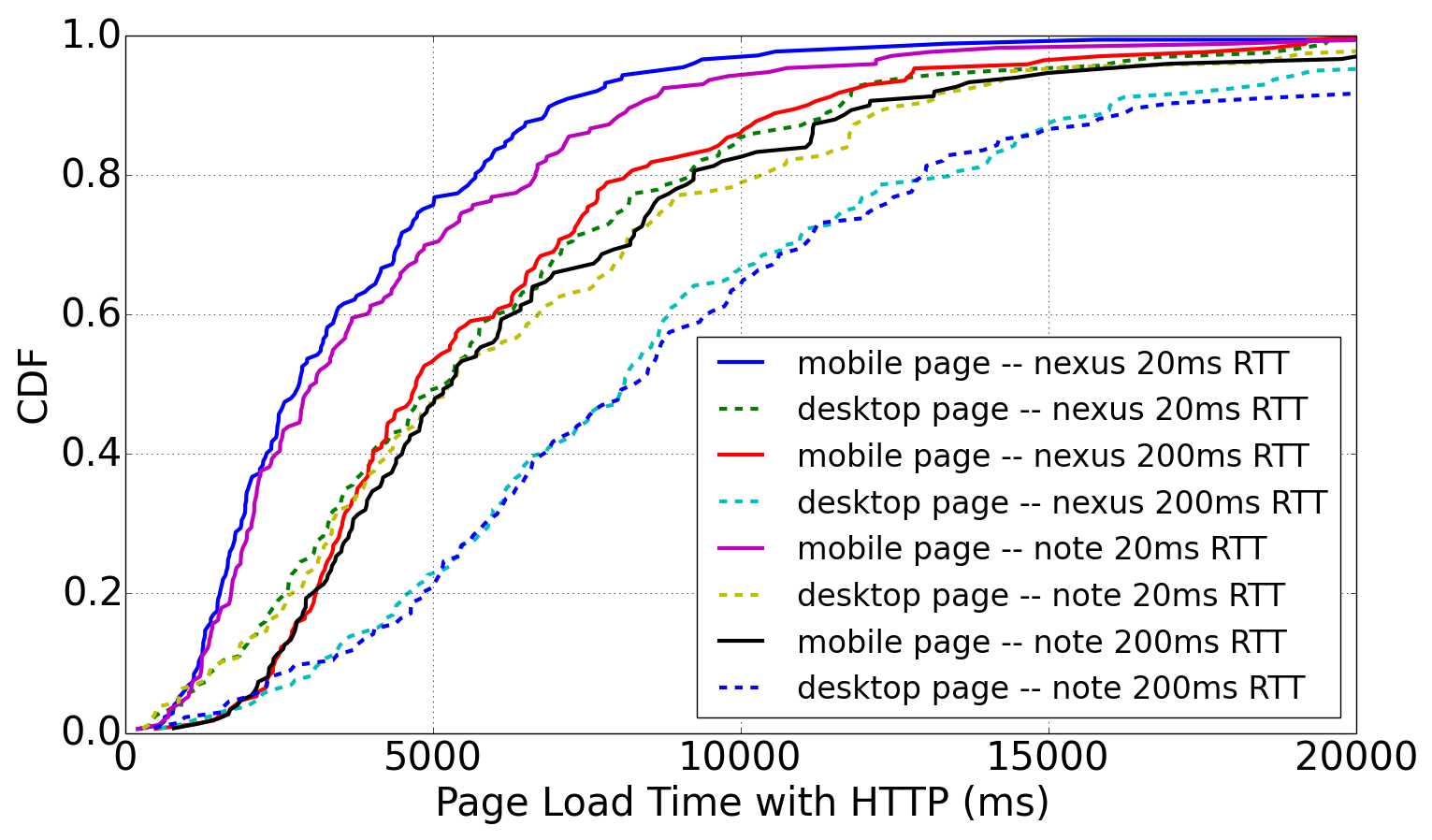}}
\subfigure[Page load time when loading pages in Nexus 6 phones and Galaxy Note 2 phones with HTTPS]{
\label{fig:device_pageload_https}
\includegraphics[width=0.48\textwidth]{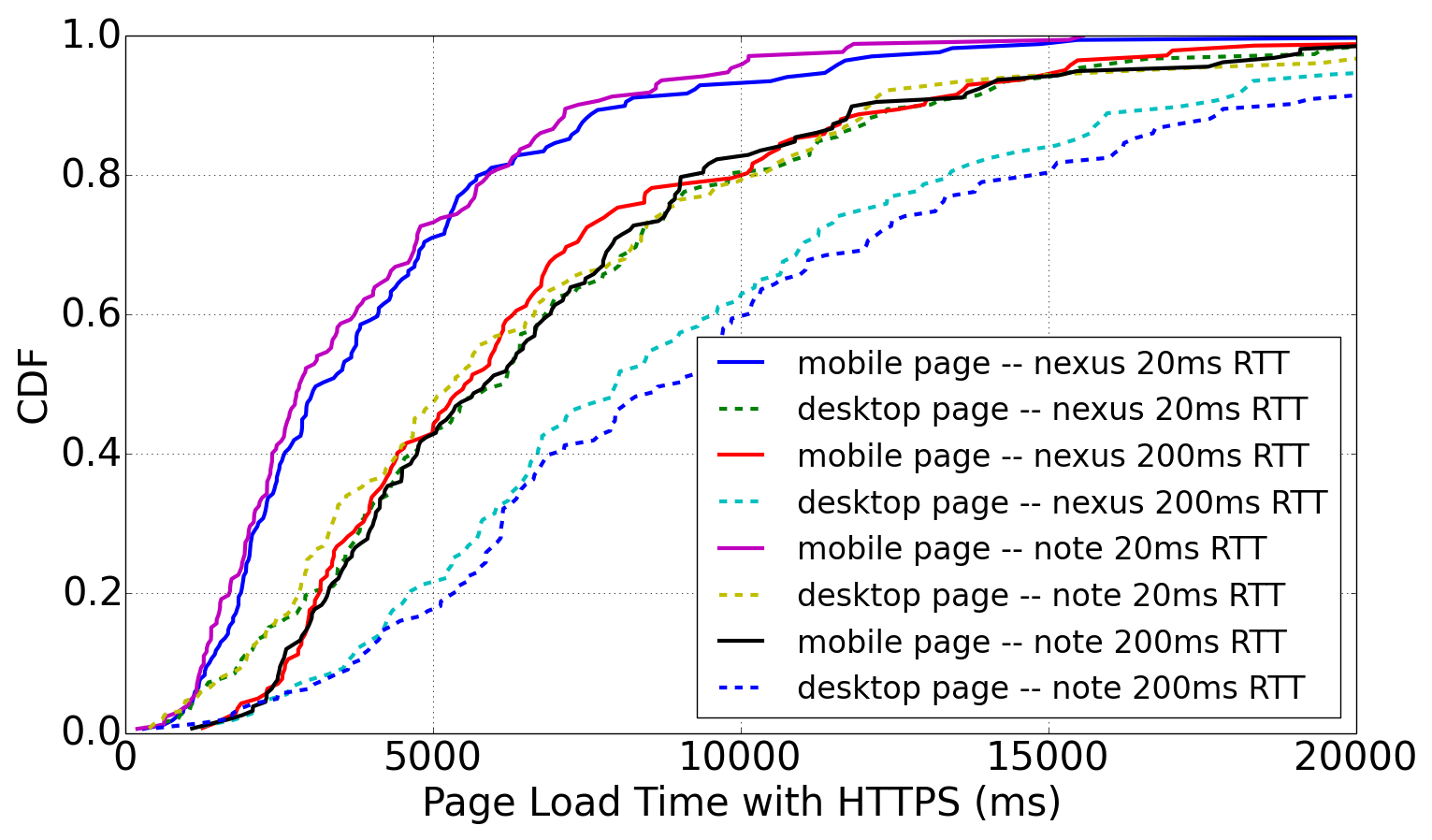}}
\subfigure[Page load time when loading pages in Nexus 6 phones and Galaxy Note 2 phones with SPDY]{
\label{fig:device_pageload_spdy}
\includegraphics[width=0.48\textwidth]{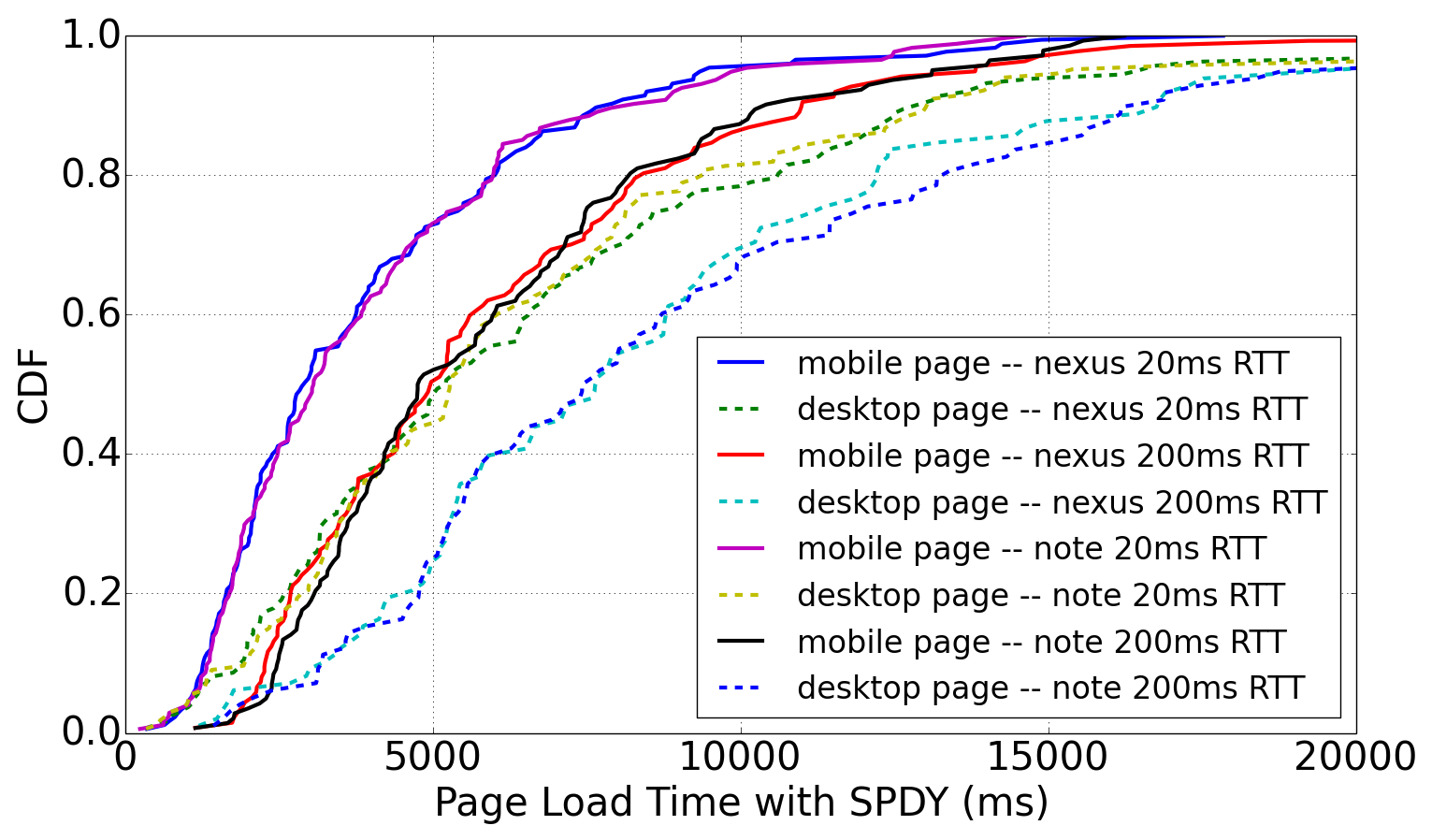}}
\subfigure[Page load time when loading pages in Nexus 6 phones and Galaxy Note 2 phones with HTTP/2]{
\label{fig:device_pageload_http2}
\includegraphics[width=0.48\textwidth]{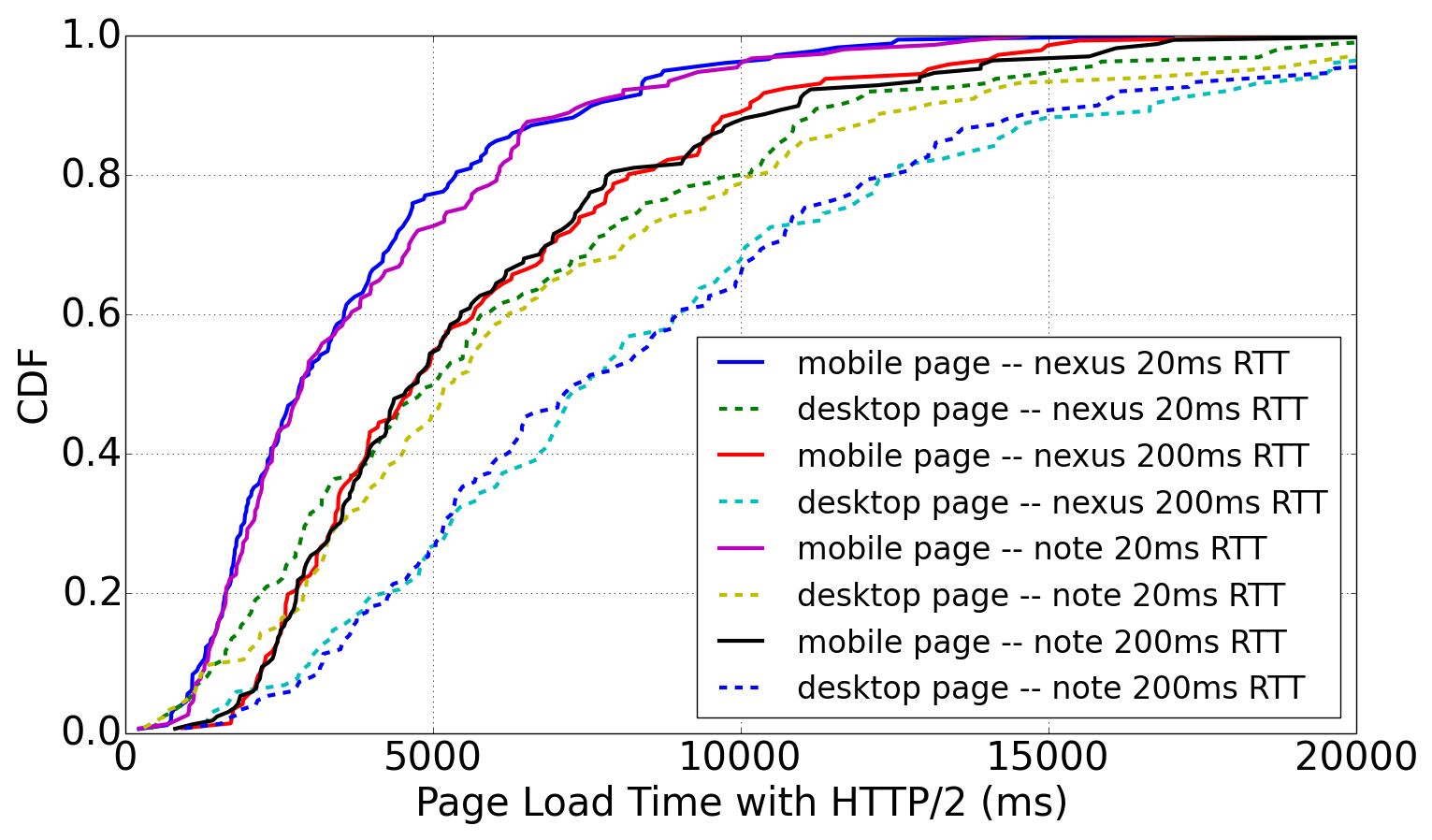}}
\caption{Page load time when loading pages in Nexus 6 phones and Galaxy Note 2 phones. Results are tested under 20ms RTT.}
\label{fig:device_pageload}
\end{figure}

Figure~\ref{fig:device_pageload} depicts the distribution of page load time when loading pages using Nexus 6 and Galaxy Note 2. We only show results under 0\% packet loss rate and 1mbps bandwidth. Figure~\ref{fig:device_pageload} shows that page load time on Galaxy Note 2 is not significantly different compared to the page load time on Nexus 6 for both desktop pages and mobile pages across all four protocols.~\cite{Zhu:MICRO15} concludes that mobile CPU performance improvements yield marginal improvements in Web browsing performance with high clock frequency. We could infer that mobile browsers are not able to take full advantage of high-performance CPU when exceeding a threshold.
% TODO 怎么细粒度点分析？还是说是网络请求占主要，所以即使计算提升了，效果也不明显

%% file: section/how_http2_help.tex
\section{How can HTTP/2 and SPDY help}

\begin{figure}[htbp]
\centering
\subfigure[PLT of HTTP/2 divided by PLT of HTTP with 20ms RTT and 1mbps bandwidth]{
\label{fig:http2_vs_http_pkl}
\includegraphics[width=0.32\textwidth]{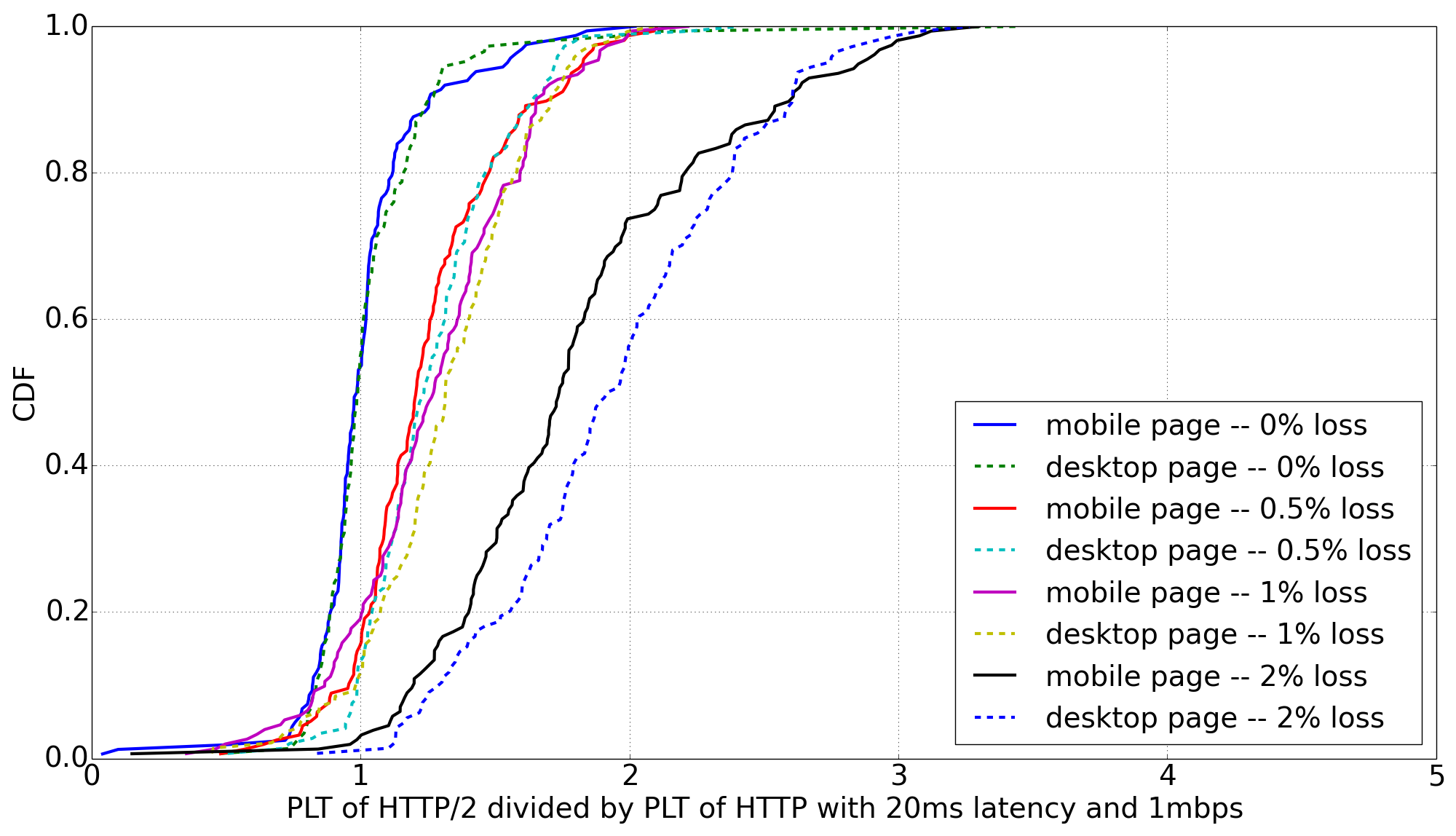}}
\subfigure[PLT of HTTP/2 divided by PLT of HTTP with 0\% packet loss rate and 1mbps bandwidth]{
\label{fig:http2_vs_http_latency}
\includegraphics[width=0.32\textwidth]{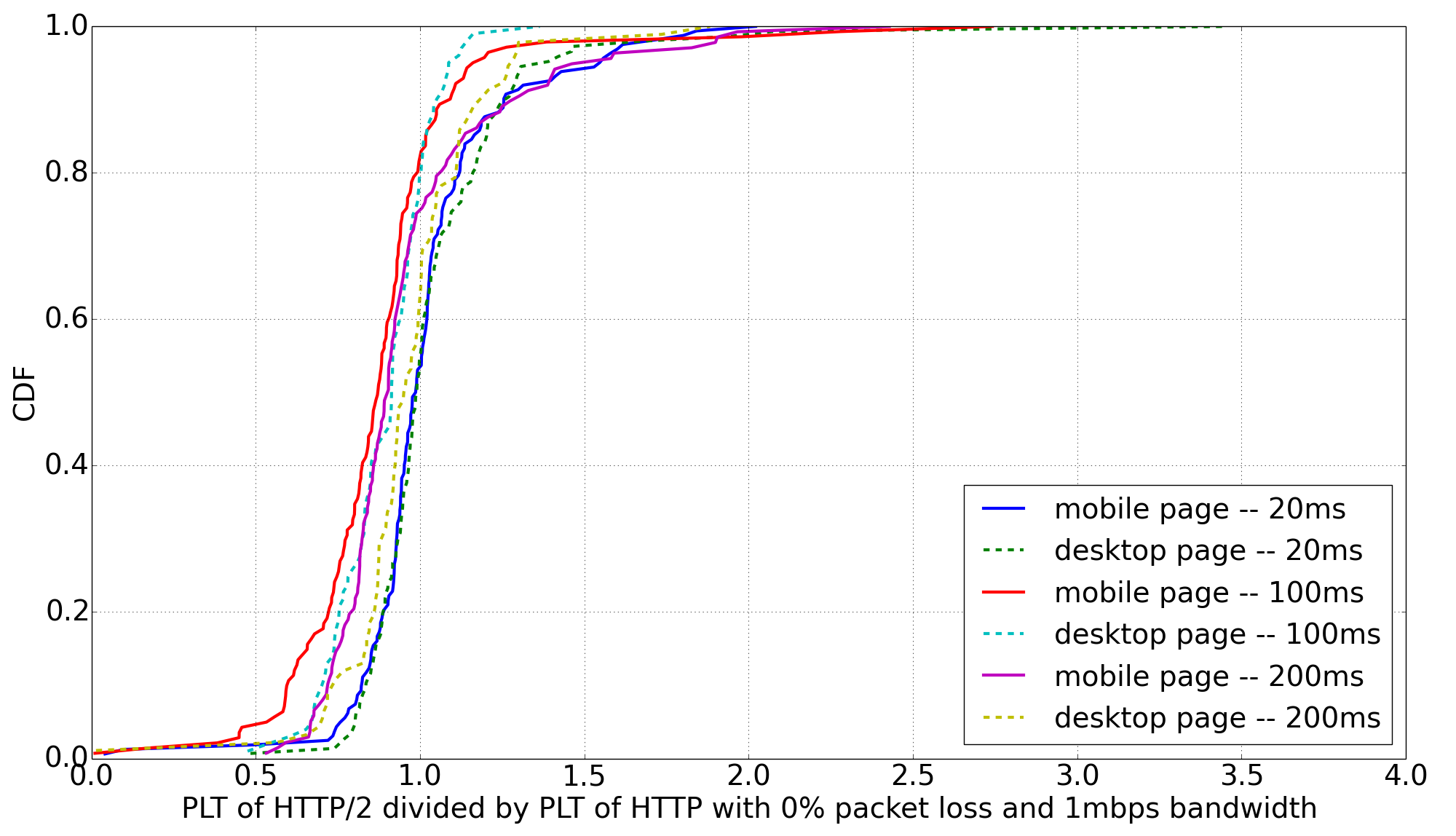}}
\subfigure[PLT of HTTP/2 divided by PLT of HTTP with 2\% packet loss rate and 1mbps bandwidth]{
\label{fig:http2_vs_http_latency_2pkl}
\includegraphics[width=0.32\textwidth]{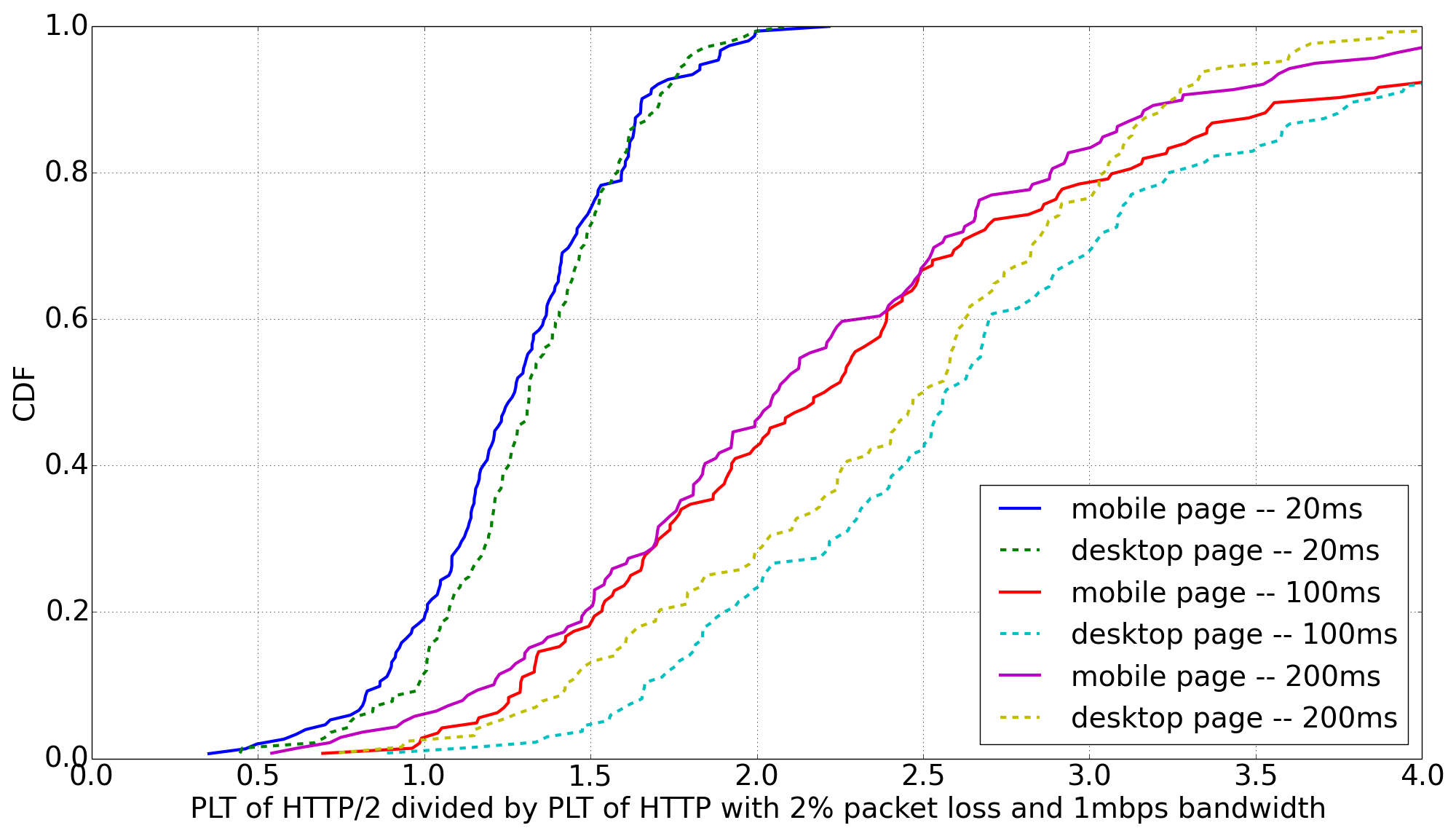}}
\caption{PLT of HTTP/2 divided by PLT of HTTP.}
\label{fig:http2_vs_http}
\end{figure}

HTTPS piggybacks HTTP entirely on top of TLS/SSL, whose end-end encryption ensures complete privacy. However, it needs one more expensive TLS/SSL handshake for each new connection, which costs extra tow Round-Trip Time. HTTPS slows down the landing page render as it takes a few hundred extra milliseconds to set up the HTTPS connection, which hinders its deployments in despite of its security.

SPDY is designed by Google to help reduce latency and bolster security. HTTP/2 is the next evolution of HTTP and is based on SPDY. HTTP/2 and SPDY attempt to outcome the shortcomings of HTTP, and focuses on performance, e.g., end-user perceived latency, network, and server resource usage. HTTP/2 and SPDY benefit from multiplexing and concurrency, stream dependencies, header compressions, and server push. After more than two years' discussion, 17 drafts, and 30 implements, IETF HTTP Working Group for publication as standards-track RFCs approved the HTTP/2 and associated HPACK specifications in February of 2015. Google have also announced to replace SPDY with HTTP/2. It is useful to deploy HTTP/2 or SPDY to speed Web pages as well as keep the security of HTTPS. SPDY shows a similar trend to HTTP/2, so we just show performance  of HTTP/2 versus HTTP under certain networks.

Figure~\ref{fig:http2_vs_http} shows the distribution of PLT of HTTP/2 divided by PLT of HTTP under different network conditions. We can find how HTTP/2 help under different conditons. Figure~\ref{fig:http2_vs_http_pkl} shows the PLT of HTTP/2 divided by PLT of HTTP when injecting different packet loss. We can find that Web page performance improvement gained by HTTP/2 decreases as packet loss rate raised. HTTP/2 reduces the PLT for 2.6\% $\sim$ 53.1\% of mobile pages, and help for 0.7\% $\sim$ 54.8\% of desktop pages. Meanwhile, we find that HTTP/2 gain more performance improvement of mobile pages compared to desktop pages, especially under high packet loss rate. Figure~\ref{fig:http2_vs_http_latency} and Figure~\ref{fig:http2_vs_http_latency_2pkl} show performance of HTTP/2 versus HTTP with different RTTs when injecting 0\% and 2\% packet loss rate, respectively. Figure~\ref{fig:http2_vs_http_latency} shows that HTTP/2 can help 53.1\% $\sim$ 82.3\% of mobile pages and 54.8\% $\sim$ 79.2\% of desktop pages. We find that HTTP/2 gain least performance improvements under low latency (20ms). However, Figure~\ref{fig:http2_vs_http_latency_2pkl} shows a reverse trend under high packet loss rate. This finding is consistent as our decision analysis, which shows that packet loss rate has more effect on performance of HTTP/2 and SPDY. Meanwhile, HTTP/2 is beaten by HTTP with at most 4\% performance improvement. Our results are consistent with prior studies~\cite{Saxce:INFCOM15, Varvello:CORR15}, which have shown that HTTP/2 only help under certain network conditions.

Overall, we need to address that HTTP/2 and SPDY do not always gain performance improvements. HTTP/2 helps under low packet loss rate, and performs better when loading mobile pages. When packet loss rate is low, HTTP/2 helps under high latency. We have discussed in previous section that high RTT favors HTTP/2 and SPDY against HTTP due to multiplexing. HTTP/2 and SPDY benefits from having a single connection and stream multiplexing. One connection per origin significantly reduces the associated overhead: fewer sockets to manage along the connection path, smaller memory footprint, better connection throughput, less time in slow-start, faster congestion and loss recovery. As the RTT goes up, new established TCP connections cost more time.

%% file: section/findings_and_implications.tex
\section{Findings and Implications}
In this paper, we want to analyze how Web pages performance vary over various parameter settings with different protocols. We conduct our measurement study of Web page performance with HTTP, HTTPS, SPDY, and HTTP/2. First, we compare these four protocols as transfer protocols with our own client and conduct a decision tree analysis to find out how those parameter settings affect these protocols` efficiency of transferring Web contents hosted in remote servers. Then, we use real browsers to load pages with these protocols to analyze how these protocols work in real browsers. We have findings as following :
 \begin{itemize}

    \item{HTTP/2 and SPDY perform worse with high packet loss. HTTP/2 and SPDY both have the feature of one TCP connection per origin. A single connection hurts under high packet loss because it aggressively reduces the congestion window compared to HTTPS which reduces the congestion window on only one of its parallel connections. When this single connection suffers from packet loss, all streams running over this unique TCP connection are negatively impacted.}
    
    \item{HTTP/2 and SPDY perform better with many objects under low loss. TCP implements congestion control by counting outstanding packets not bytes. Thus, sending a few small objects with HTTP will promptly use up the congestion window, though outstanding bytes are far below the window limit. In contrast, A single HTTP/2 connection can contain multiple concurrently-open streams, with either endpoint interleaving frames from multiple streams.}
    
    \item{The complicated dependencies and computation in real browsers may blur the bright spot of HTTP/2 and SPDY. Loading a Web page is not as simple as fetching all resources in parallel. Loading a Web page in modern browsers is a complex work, including parsing HTML, parsing JavaScript/CSS, and interpreting JavaScript/CSS etc.}
    
    \item{The mobile pages perform better than respective desktop pages. Mobile pages are dedicated for mobile devices with limited capabilities of computation and smaller screen, and may be optimized with less resources and smaller objects. Mobile pages under bad network conditions may even perform better than respective desktop pages under good network conditions.}
    
    \item{PLT on Galaxy Note 2 is not significantly different compared to the page load time on Nexus 6 for both desktop pages and mobile pages across all four protocols. We could infer that 1) mobile browsers are not able to take full advantage of high-performance CPU when exceeding a threshold; 2) network activities may dominate in the page load process against computation activities, and the improvement of computation just have marginal effect on whole page load process.}
        
 \end{itemize}
 
 In our experiments, we do find that the benefits from HTTP/2 when loading real Web pages using mobile browsers, especially under certain network conditions. However, HTTP/2 hurts under other network conditions, and the performance could be affected by the characteristics of Web pages. We give some implications as shown in Table~\ref{tab:findings and implications}. For website developers, they should adjust their expectations that it may not speed their Web pages. It may be more practical to optimize their websites to reduce Web object sizes, improve cache configures etc. Of course, we should not be too negative, since mobile devices are becoming more and more powerful and related technologies of HTTP/2 is advancing. If developers want to deploy HTTP/2, they should carefully consider where their customers will visit their Web pages. It is better to switch the protocols according the network conditions when customers visit the Web pages. For example, client developers can force the client to fetch resources using HTTP/2 when devices are connected to a good WiFi and using HTTP when   connected to 2G/3G.

 \begin{table*}[htbp]\normalsize
\newcommand{\tabincell}[2]{\begin{tabular}{@{}#1@{}}#2\end{tabular}}
  \centering
  \caption{Summary of Findings and Implications}\label{tab:findings and implications}
  \begin{tabularx}
{\linewidth}{|X|X|}
\hline
    \rowcolor{mygray}\bfseries Findings & \bfseries Implications \\
%\caption{Our major findings and implications}\label{tab:findings and implications}
%% Some packages, such as MDW tools, offer better commands for making tables
%% than the plain LaTeX2e tabular which is used here.
%\begin{tabular}{| p{8cm}| p{8cm}|}
%\hline
%\rowcolor{lightgray} {\centering} Findings & Implications\\
\hline
HTTP/2 and SPDY could both decrease and increase page load times under certain network conditions and page characteristics. For example, HTTP/2 and SPDY helps Web pages with many Web objects when packet loss rate is low.
&
Developers should consider the adoption of HTTP/2 and SPDY thoroughly. It will not speed their websites with few Web objects. It is necessary to characterize websites before turning to HTTP/2 and SPDY.\\
\hline
Different protocols perform variously under different parameter settings. For example, HTTP/2 and SPDY perform better with many objects under low loss, but perform worse when packet loss rate is high.
&
Developers should carefully consider where their customers will visit their Web pages. It is better to switch the protocols according the network conditions when customers visit the Web pages. For example, client developers can force the client to fetch resources using HTTP/2 when devices are connected to a good WiFi and using HTTP when   connected to 2G/3G.
\\
\hline
Mobile Web pages have much better performance than the respective desktop pages. Mobile pages under bad network conditions may even perform better than respective desktop pages under good network conditions.
&
More and more customers are used to visiting Web pages using mobile devices, which have smaller screen, limited capabilities of computation and power. Developers should optimize their Web pages to improve user experience on mobile devices. For customers, they should visit mobile page instead of desktop page when using mobile devices.
\\
\hline
More powerful devices may not speed the page load. Loading Web page in browser is a complicated process, consisting of network activities and computation activities. Powerful CPU may speed the computation activities, but may just gain marginal improvements as a whole.
&
Developers should not just focus on the optimization of computation activities in a Web page, but also the optimization of network activities. They could try HTTP/2 to speed page load. Customers should not rely on more powerful devices to speed Web pages visiting completely.
\\
\hline
\end{tabularx}
\end{table*} 

%% file: section/relative_work.tex
\section{Related Work}
Web service publishers aim to provide competitive services, and utilize all kinds of technologies to improve services' performance. In this paper, we focus on how HTTP/2 could reduce users perceived latency, which is a important factor of QoS. HTTP/2 derives from SPDY and aims to improve the performance of HTTP, so we focus on prior studies on improvement of HTTP and works about HTTP/2 and SPDY in this section.

\textbf{Quality of Service (QoS):} QoS is always attractive research topic of Web service. Ma et al.~\cite{Ma:SCC13} investigate how to measure Web service QoS precisely from both subjective and objective aspects. Nacer et al.~\cite{Ahmed-Nacer:SCC15} and Chen et al.~\cite{Chen:SCC15} present service selection and service recommendation, respectively. Elshater et al.~\cite{Elshater:SCC15} utilizes design pattern to improve performance in terms of average response time and throughput. Liu et al.~\cite{liuTSC2009}  alleviates the consumers from time-consuming discovery tasks. However, the performance of services could be easily affected by network conditions and complexity of services. It is important to speed the delivery of response content to services consumers more than services processing. 

\textbf{HTTP:} Radhakrishnan et al.~\cite{Radhakrishnan:CONEXT11} describes the TCP Fast Open protocol, a new mechanism that enables data exchange during TCP`s initial handshake. Flach et al.~\cite{Flach:SIGCOMM13} presents the design of novel loss recovery mechanisms for TCP that judiciously use redundant transmissions to minimize timeout-driven recovery.

\textbf{SPDY:} Erman et al.~\cite{Erman:CONEXT13} finds that SPDY performed poorly while interacting with radios due to a large body of unnecessary retransmissions. Wang et al.~\cite{Wang:NSDI14} swipe a complete parameter space including network parameters, TCP settings, and Web page characteristics to learn which factors affect the performance of SPDY. Thus, they also dive into the analysis of the effect of dependencies of network activities and computation activities in real browsers. However, we are more curious how HTTP/2 performs in mobile devices with limited capacities and poor network conditions. El-khatib et al.~\cite{El-khatibTW:IFIP14} also analyze the effect of network and infrastructural variables on SPDY's performance. Prior works on SPDY do encourage us to reveal the mysterious mask of HTTP/2. 

\textbf{HTTP/2:} Chowdhury et al.~\cite{Chowdhury:PEERJ15} focus on the energy efficiency of HTTP/2, and shows that HTTP/2 exhibits better performance as the RTT goes up. However, we are more curious if HTTP/2 could reduce page load time. Hugues et al.~\cite{Saxce:INFCOM15} evaluate the influence of HTTP/2` new features on the Web performance. They notice generally speedier page load times with HTTP/2, thanks to the multiplexing and compression features enabled by the protocol. The limitation is that they only test 15 websites. We test the Alexa Top 200 websites with both mobile version and desktop version in our experiment. Zarifis et al.~\cite{Zarifis:PAM16} develop a model to compare the PLT of HTTP/1.1 and HTTP/2 from the resource timing data for a single HTTP/1.1 page view. Varvello et al.~\cite{Varvello:PAM16} present a measurement platform to monitor both adoption and performance of HTTP/2. They find that developers do not reconstruct their Web pages when turning to HTTP/2 and 80\% of websites supporting HTTP/2 experience a decrease in page load time compared With HTTP/1.1 and the decrease grows in mobile networks. Although HTTP/2 announces to provide better performance, real adoptions do not gain much performance improvements and developers may still be in doubt. 

%% file: section/conclusion.tex
\section{Conclusion}

In this paper, we want to analyze how Web pages performance vary over various parameter settings, including RTT, bandwidth, loss rate, number of objects on a page, and objects sizes, with different protocols. We conduct our measurement study of Web page performance with HTTP, HTTPS, SPDY, and HTTP/2. First, we compare these four protocols as transfer protocols with our own client and conduct a decision tree analysis to find out how those parameter settings affect these protocols` efficiency of transferring Web contents hosted in remote servers. We use Linux Traffic Control (TC) to emulate different network to ensure browsers load pages under consistent network conditions for each protocol. We find that HTTP/2 and SPDY perform worse when packet loss rate is high, but help with many objects when packet loss rate is low. Then, we use real browsers to load pages with these protocols to analyze how these protocols work in real browsers. We clone the landing pages of the Alexa top 200 websites that have corresponding mobile version into our local server and convert all external links to local links. Overall, we collect 400 Web pages, consisting of 200 desktop pages and 200 corresponding mobile pages. We find that mobile pages have better PLTs that respective desktop pages, and HTTP/2 helps under low packet loss rate, and performs better when loading mobile pages. Meanwhile, we conduct our previous controlled experiments on a older device to observer if different devices could affect Web page performance. We find similar performance for all four protocols under different network conditions. We infer that mobile browsers are not able to take full advantage of high-performance CPU when exceeding a threshold.